\documentclass[conference]{IEEEtran}
\usepackage{cite}
\usepackage{amsmath,amssymb,amsfonts}
\usepackage{algorithm}
\usepackage{graphicx}
\usepackage{textcomp}
\usepackage{xcolor}
\usepackage{algpseudocode}
\usepackage{paralist}
\graphicspath{{images/}}
\usepackage{subfigure}
\usepackage{url}
\usepackage{comment}
\usepackage{wrapfig}
\usepackage{todonotes}

\algdef{SE}[EVENT]{Event}{EndEvent}[1]{\textbf{upon}\ #1\
\algorithmicdo}{\algorithmicend\ \textbf{}}
\def\NoNumber#1{{\def\alglinenumber##1{}\State #1}\addtocounter{ALG@line}{-1}}

\def\BibTeX{{\rm B\kern-.05em{\sc i\kern-.025em b}\kern-.08em
    T\kern-.1667em\lower.7ex\hbox{E}\kern-.125emX}}
\begin{document}

\title{SSS: Scalable Key-Value Store with External Consistent and Abort-free Read-only Transactions}

\author{\IEEEauthorblockN{Masoomeh Javidi Kishi, Sebastiano Peluso\textsection, Hank Korth, Roberto Palmieri}
\IEEEauthorblockA{\textit{Lehigh University, Virginia Tech\textsection}\\
Bethlehem, PA, USA \\
\{maj717;hfk2;palmieri\}@lehigh.edu; \textsection peluso@vt.edu}
}

\maketitle

\begin{abstract}
We present SSS, a scalable transactional key-value store deploying a novel distributed concurrency control that provides external consistency for all transactions, never aborts read-only transactions due to concurrency, all without specialized hardware. SSS ensures the above properties without any centralized source of synchronization. SSS's concurrency control uses a combination of vector clocks and a new technique, called snapshot-queuing, to establish a single transaction serialization order that matches the order of transaction completion observed by clients. We compare SSS against high performance key-value stores, Walter, ROCOCO, and a two-phase commit baseline. SSS outperforms 2PC-baseline by as much as 7x using 20 nodes; and ROCOCO by as much as 2.2x with long read-only transactions using 15 nodes.
\end{abstract}


\section{Introduction}

A distributed transactional system that ensures a strong level of consistency greatly simplifies programmer responsibility while developing applications.
A strong level of consistency that clients interacting with a transactional system often desire is referred to \textit{external consistency}~\cite{spanner,strict,exconsistency}.

Roughly, under external consistency a distributed system behaves as if all transactions were executed sequentially, all clients observed the same unique order of transactions completion (also named \textit{external schedule} in~\cite{exconsistency}), in which every read operation returns the value written by the previous write operation.
By relying on the definition of external consistency, a transaction
terminates when its execution is returned to its client;
therefore the order defined by transaction client returns matches the transaction serialization order.

The latter property carries one great advantage:
if clients communicate with each other outside the system, they cannot be confused about the possible mismatch between transaction order they observe and the transaction serialization order provided by the concurrency control inside the system.
Simply, if a transaction is returned to its client, the serialization order of that transaction will be after any other transaction returned earlier, and before any transaction that will return subsequently. Importantly, this property holds regardless of when transactions start their execution.


To picture the value of external consistency,
consider an online document sharing service and two clients, $C_1$ connected to server $N_1$ and $C_2$ connected to another server $N_2$, whose goal is to synchronize the same document $D$. $C_1$ modifies $D$ and starts its synchronization concurrently with $C_2$. Because the underlying distributed system that implements the service is asynchronous, it is plausible for $C_1$ to observe the completion of the synchronization before $C_2$ (e.g., $C_1$ and $C_2$ are handled by two different nodes with different speed). As soon as $C_1$ received the notification, it informs $C_2$ that its modifications are permanent. At this point $C_2$ observes the completion of its synchronization operation and its expectation is to observe $C_1$'s modification on $D$ since $C_1$ completed before $C_2$.

Only if the service is external consistent, then the expectation is met (i.e., $C_2$ observes the modification of $C_1$); otherwise the possible outcomes include the case where $C_2$ does not observe $C_1$, which might confuse $C_2$. Note that if the service provides Serializable~\cite{serializability} operations, $C_2$ will not be guaranteed to observe the outcome of $C_1$.

In this paper we present \textit{SSS}, a key-value store that implements a novel distributed concurrency control providing external consistency and assuming off-the-shelf hardware.
Two features enable high performance and scalability in SSS, especially in read-dominated workloads:
\begin{itemize}
\item SSS supports read-only transactions that never abort due to concurrency, therefore the return value of all their read operations should be consistent at the time the operation is issued.
We name them as \textit{abort-free} hereafter.
This property is very appealing because many real-world applications produce significant read-only workload~\cite{DBLP:conf/sigmetrics/AtikogluXFJP12}.

\item SSS is designed to support a general (partial) replication scheme where keys are allowed to be maintained by any node of the system without predefined partitioning schemes (e.g., sharding~\cite{tapir,scatter}). To favor scalability, SSS does not rely on ordering communication primitives, such as Total Order Broadcast or Multicast~\cite{DBLP:journals/csur/DefagoSU04}.

\end{itemize} 

The core components that make the above properties possible in SSS are the following:

\begin{compactitem}

\item SSS uses a vector clock-based technique to track dependent events originated on different nodes. This technique is similar to the one used by existing distributed transactional systems, such as Walter~\cite{walter} and GMU~\cite{gmu12}, and allows SSS to track events without a global source of synchronization.

\item SSS uses a new technique, which we name \textit{snapshot-queuing}, that works as follows. Each key is associated with a \textit{snapshot-queue}.
Only transaction that will surely commit
are inserted into the snapshot-queues of the their accessed keys in order to leave a trace of their existence to other concurrent transactions. Read-only transactions are inserted into their read keys' snapshot-queues at read time, while update transactions into their modified keys' snapshot-queues after the commit decision is reached.

A transaction in a snapshot-queue is inserted along with a scalar value, called \textit{insertion-snapshot}.
This value represents the latest snapshot visible by the transaction on the node storing the accessed key,
at the time the transaction is added to the snapshot queue. SSS concurrency control orders transactions with lesser insertion-snapshot before conflicting transactions with higher insertion-snapshot in the external schedule.

\end{compactitem}
SSS uses snapshot-queues to propagate established serialization orders among concurrent transactions as follows.

If a read-only transaction $T_R$ reads a key $x$ subsequently modified by a concurrent committed transaction $T_W$, $x$'s snapshot-queue is the medium to record the existence of an established serialization order between $T_R$ and $T_W$. With that, any other concurrent transaction accessing $x$ can see this established order and define its serialization accordingly.

In addition, $T_W$'s client response is delayed until $T_R$ completes its execution. This delay is needed so that update transactions can be serialized along with read-only transactions in a unique order where reads always return values written by the last update transaction returned to its client.

Failing in delaying $T_W$'s response would result in a discrepancy between the external order and the transaction serialization order. In fact, the external order would show $T_W$ returning earlier than $T_R$ but $T_R$ is serialized before $T_W$.

For non-conflicting update transactions that have dependencies with concurrent read-only transactions accessing common keys, since these transactions are aware of each other through the snapshot-queues of accessed keys, SSS prevents read-only transactions to observe those update transactions in different orders.







On the flip side, delaying update transactions might have a domino effect on limiting the level of concurrency in the system, which might lead to poor performance. The snapshot-queue technique prevents that: it permits a transaction that is in a snapshot-queue
to expose its written keys to other transactions while it is waiting for the completion of the concurrent read-only transaction(s) holding it. This feature enables progress of subsequent conflicting transactions, hence retaining the high throughput of the system.

Update transactions in SSS are serialized along with read-only transactions.
They always read the latest version of a key and buffer write operations. Validation is performed at commit time to abort if some read key has been overwritten meanwhile. The Two-Phase Commit protocol (2PC)~\cite{strict,gmu12, walter, Cockroach, carousel} is used to atomically lock and install written keys. These keys are externally visible when no concurrent read-only transactions caused the update transaction to wait due to snapshot-queuing, if any.



We implement SSS in Java and compared against two recent key-value stores, Walter~\cite{walter} and ROCOCO~\cite{rococo}, and one baseline where all transactions, including read-only, validate read keys and use 2PC to commit~\cite{serializability}. We name this competitor 2PC-baseline.
Overall, SSS is up to 7$\times$ faster than 2PC-baseline and up to 2.2$\times$ faster than ROCOCO under read-dominated workloads and long (i.e., 16 read keys) read-only transactions.
Also, when the percentage of read-only transactions is dominant and the node count is high, SSS is only 18\% slower than Walter, which provides a weaker isolation level than external consistency and even serializability.
When compared to the overall update transaction latency, in our experiments we assessed in less than 28\% the average waiting time introduced by SSS due to the snapshot-queuing.

\section{System Model \& Assumptions}

SSS assumes a system as a set of nodes that do not share either memory or a global clock. Nodes communicate through message passing and reliable asynchronous channels, meaning messages are guaranteed to be eventually delivered unless a crash happens at the sender or receiver node. There is no assumption on the speed and on the level of synchrony among nodes.
We consider the classic crash-stop failure model: sites may fail by crashing, but do not behave maliciously. A site that never crashes is correct; otherwise it is faulty.
Clients are assumed to be colocated with nodes in the system; this way a client is immediately notified of a transaction's commit or abort outcome, without additional delay.
Clients are allowed to interact with each other through channels that are not provided by the system's APIs.

\underline{Data Organization}.
Every node $N_{i}$ maintains shared objects (or keys) adhering to the key-value model~\cite{gmu12}. Multiple versions are kept for each key. Each version
stores the value and the commit vector clock of the transaction that produced the version.
SSS does not make any assumption on the data clustering policy; simply every shared key can be stored in one or more nodes, depending upon the chosen replication degree. For object reachability, we assume the existence of a local look-up function that matches keys with nodes.

\underline{Transaction execution}.
We model transactions as a sequence of read and write operations on shared keys, preceded by a begin, and followed by a commit or abort operation. A client begins a transaction on the colocated node and the transactions can read/write data belonging to any node; no a-priori knowledge on the accessed keys is assumed.
SSS's concurrency control ensures the ACID properties and targets applications with a degree of data replication.

Every transaction starts with a client submitting it to the system,
and finishes its execution informing the client about its final outcome: \textit{commit} or \textit{abort}.
Transactions that do not execute any write operation are called read-only, otherwise they are update transactions.
SSS requires programmer to identify whether a transaction is update or read-only.





\section{SSS Concurrency Control}
\label{sec:over}

In this section we describe the SSS concurrency control, followed by two execution examples.

\subsection{Metadata}

\underline{\textit{Transaction vector clocks}}. In SSS a transaction $T$ holds two vector clocks, whose size is equal to the number of nodes in the system. One represents its actual dependencies with transactions on other nodes, called \texttt{T.VC}; the other records the nodes where the transaction read from, called \texttt{T.hasRead}.

\texttt{T.VC} represents a version visibility bound for $T$.
Once a transaction begins in node $N_{i}$, it assigns the vector clock of the latest committed transaction in $N_{i}$
to its own \texttt{T.VC}.
Every time $T$ reads from a node $N_{j}$ for the first time during its execution, \texttt{T.VC} is modified based on the latest committed vector clock visible by $T$ on $N_{j}$.
After that, \texttt{T.hasRead[j]} is set to true ($\top$).


\underline{\textit{Transaction read-set and write-set}}. Every transaction holds two private buffers. One is $rs$ (or \textit{read-set}), which stores the keys read by the transaction during its execution, along with their value.
The other buffer is $ws$ (or \textit{write-set}), which contains the keys the transaction wrote, along with their value.

\underline{\textit{Snapshot-queue}}. A fundamental component allowing SSS to establish a unique external schedule is the snapshot-queuing technique. With that, each key is associated with an ordered queue (\texttt{SQueue}) containing: read-only transactions that
read that key; and update transactions that wrote that key while a read-only transaction was reading it.

Entries in a snapshot-queue (\texttt{SQueue}) are in the form of tuples.
Each tuple contains: transaction identifier $T.id$, the \textit{insertion-snapshot}, and transaction type (read-only or update). The \textit{insertion-snapshot} for a transaction $T$ enqueued on some node $N_i$'s snapshot-queue is the value of $T$'s vector clock in position $i^{th}$ at the time $T$ is inserted in the snapshot-queue.
Transactions in a snapshot-queue are ordered according to their insertion-snapshot.

A snapshot-queue contains only transactions that will commit; in fact, besides read-only transactions that are abort-free, update transactions are inserted in the snapshot-queue only after their commit decision has been reached.

\underline{\textit{Transaction transitive anti-dependencies set}}. An update transaction  maintains a list of snapshot-queue entries, named \texttt{T.PropagatedSet}, which is populated during the transaction's read operations. This set serves the purpose of propagating anti-dependencies previously observed by conflicting update transactions.

\underline{\textit{Node's vector clock}}. Each node $N_{i}$ is associated with a vector clock, called \texttt{NodeVC}.
The $i^{th}$ entry of \texttt{NodeVC} is incremented when $N_{i}$ is involved in the commit phase of a transaction that writes some key replicated by $N_i$.
The value of $j^{th}$ entry of \texttt{NodeVC} in $N_{i}$ is the value of the $j^{th}$ entry of \texttt{NodeVC} in $N_{j}$
at the latest time $N_{i}$ and $N_{j}$ cooperated in the commit phase of a transaction.

\underline{\textit{Commit repositories}}. \texttt{CommitQ} is an ordered queue, one per node, which is used by SSS to ensure that non-conflicting transactions are ordered in the same way on the nodes where they commit.
\texttt{CommitQ} stores tuples $<$$T$, $vc$, $s$$>$ with the following semantics.
When an update transaction $T$, with commit vector clock $vc$, enters its commit phase, it is firstly added to the \texttt{CommitQ} of the nodes participating in its commit phase with its status $s$ set as \textit{pending}.

When the outcome of the transaction commit phase is decided, the status of the transaction is changed to \textit{ready}. A ready transaction inside the \texttt{CommitQ} is assigned with a final vector clock produced during the commit phase. In each node $N_{i}$, transactions are ordered in the \texttt{CommitQ} according to the $i^{th}$ entry of the vector clock ($VC[i]$). This allows them to be committed in $N_{i}$ with the order given by $VC[i]$. 

When $T$ commits, it is deleted from \texttt{CommitQ} and its $vc$ is added to a per node repository, named \texttt{NLog}. We identify the most recent $vc$ in the \texttt{NLog} as \texttt{NLog.mostRecentVC}.

Overall, the presence of additional metadata to be transferred over the network might appear as a barrier to achieve high performance. To alleviate these costs we adopt metadata compression. In addition, while acknowledging that the size of vector clocks grows linearly with the system size, there are existing orthogonal solutions to increase the granularity of such a synchronization to retain efficiency~\cite{VCOptimized,VCOptimized1}




\subsection{Execution of Update transactions}
\label{update-tx}


Update transactions in SSS implements lazy update~\cite{DBLP:conf/sosp/TuZKLM13}, meaning their written keys are not immediately visible and accessible at the time of the write operation, but they are logged into the transactions write-set and become visible only at commit time. In addition, transactions record the information associated with each read key into their read-set.

Read operations of update transactions in SSS simply return the most recent version of their requested keys
(Lines~\ref{vis24}-\ref{vis26} of Algorithm~\ref{visibility}). At commit time, validation is used to verify that all the read versions have not been overwritten.

An update transaction that completes all its operations and commits cannot inform its client if it observes anti-dependency with one or more read-only transactions. In order to capture this waiting stage, we introduce the following phases to finalize an update transaction (Figure~\ref{fig1} pictures them in a running example).


\underline{\textit{Internal Commit}}. 
When an update transaction successfully completes its commit phase, we say that it commits internally. In this stage, the keys written by the transactions are visible to other transactions, but its client has not been informed yet about the transaction completion. Algorithms~\ref{internal1} and~\ref{internal2} show the steps taken by SSS to commit a transaction internally.

SSS relies on the Two-Phase Commit protocol (2PC) to internally commit update transactions. 
The node that carries the execution of a transaction $T$, known as its coordinator, initiates 2PC issuing the prepare phase, in which it contacts all nodes storing keys in the read-set and write-set. When a participant node $N_{i}$ receives a prepare message for $T$, all keys read/written by $T$ and stored by $N_{i}$ are locked. If the locking acquisition succeeds, all keys read by $T$ and stored by $N_{i}$ are validated by checking if the latest version of a key matches the read one (Lines~\ref{l26}-\ref{l32} Algorithm~\ref{internal1}). If successful, $N_{i}$ replies to $T$'s coordinator with a \texttt{Vote} message, along with a proposed commit vector clock. This vector clock is equal to $N_{i}$'s $NodeVC$ where $NodeVC[i]$ has been incremented. Finally, $T$ is inserted into $N_{i}$'s \texttt{CommitQ} with its $T.VC$.

\begin{algorithm}[h]
\caption{Internal Commit by Transaction T in node $N_{i}$}
\label{internal1}
\begin{algorithmic}[1]
\scriptsize
\Event {boolean Commit(Transaction T)}
\If{(T.ws=$\phi $)}
\For{($k \in T.rs$)}
\State \textbf{Send} \textit{Remove}$[T]$  \textbf{to all} replicas(k)
\EndFor
\State $T.outcome \leftarrow true$
\State \Return {$T.outcome$}
\EndIf
\State $commitVC \leftarrow T.VC$
\State $T.outcome \leftarrow true$
\State \textbf{send} \textit{Prepare}$[T]$ \textbf{to all}  ${N_j} \in replicas(T.rs\cup T.ws)\cup {N_i}$
\ForAll {(${N_j}\in replicas(T.rs \cup T.ws) \cup {N_i}$)} \label{int1-11}
\State \textbf{wait receive} \textit{Vote}$[T.id, VC_{j}, res]$ from ${N_j}$ \textbf{or timeout}
\If{($res =  false  \vee timeout$)}
\State $T.outcome \leftarrow  false $
\State break;
\Else
\State $commitVC \leftarrow \max (commitVC,V{C_j})$  \label{l17}
\EndIf
\EndFor
\State $xactVN \leftarrow \max \{ commitVC[w]:{N_w} \in replicas(T.ws) \} $
\ForAll{(${N_j} \in replicas(T.ws)$)}
\State $commitVC[j] \leftarrow xactVN$
\EndFor  \label{int1_23}
\State \textbf{send} \textit{Decide}$[T,commitVC,outcome]$ to \textbf{all} ${N_j} \in replicas($
\Statex \hspace{0.4cm} $T.rs \cup T.ws) \cup {N_i}$

\EndEvent
\Statex
\State boolean validate(Set rs, VC T.VC)   \label{l26}
\ForAll {($k \in rs$)}
\If {($k.last.vid[i] > T.VC[i]$)}
\State \Return{$ false $}
\EndIf
\EndFor
\State \Return{$true$}    \label{l32}
\end{algorithmic}
\end{algorithm}

After receiving each successful \texttt{Vote}, $T$'s coordinator updates $T.VC$ by computing the maximum per entry (Line~\ref{l17} of Algorithm~\ref{internal1}).
This update makes $T$ able to include the causal dependencies of the latest committed transactions in all 2PC participants.
After receiving all \texttt{Vote} messages, the coordinator determines the final commit vector clock for $T$ as in (Lines~\ref{l17}-\ref{int1_23} of Algorithm~\ref{internal1}), and sends it along with the 2PC \texttt{Decide} message.

Lines~\ref{l16}-\ref{l29} of Algorithm~\ref{internal2} shows how 2PC participants handle the \texttt{Decide} message. When $N_i$ receives \texttt{Decide} for transaction $T$,
$N_i$'s $NodeVC$ is updated by computing the maximum with $T.VC$.
Importantly, at this stage the order of $T$ in the $CommitQ$ of $N_{i}$ might change because the final commit vector clock of $T$ has been just defined, and it might be different from the one used during the 2PC prepare phase when $T$ has been added to $CommitQ$.

In Algorithm~\ref{internal2} Lines~\ref{l229}-\ref{l236}, when transaction $T$ becomes the top standing of $N_i$'s $CommitQ$, the internal commit of $T$ is completed by inserting its commit vector clock into the $NLog$ and removing $T$ from $CommitQ$. When transaction's vector clock is inserted into the node's \texttt{NLog}, its written keys become accessible by other transactions.
At this stage, $T$'s client has not been informed yet about $T$'s internal commit.

\underline{\textit{Pre-Commit.}}
An internally committed transaction spontaneously enters the Pre-Commit phase after that. Algorithm~\ref{finalize} shows detail of Pre-commit phase. At this stage, $T$ evaluates if it should hold the reply to its client
depending upon the content of the snapshot-queues of its written keys. If so, $T$ will be inserted into the snapshot-queue of its written keys in $N_i$ with $T.VC[i]$ as \textit{insertion-snapshot}.

If at least one read-only transaction ($T_{ro}$) with a lesser \textit{insertion-snapshot} is found in any snapshot-queue $SQueue$ of $T$'s written keys, it means that $T_{ro}$ read that key before $T_{w}$ internally committed, therefore a write-after-read dependency between $T_{ro}$ and $T_{w}$ is established. In this case, $T$ is inserted into $SQueue$ until $T_{ro}$ returns to its client. With the anti-dependency, the transaction serialization order has been established with $T_{ro}$ preceding $T_{w}$, therefore informing immediately $T_{w}$'s client about $T_{w}$'s completion would expose an external order where $T_{w}$ is before $T_{ro}$, which violates external consistency (and therefore clients expectations).

\begin{algorithm}[h]
\caption{Internal Commit by Transaction T in node $N_{i}$}
\label{internal2}
\begin{algorithmic}[1]
\scriptsize
\Event{\textbf{receive} \textit{Prepare}$[TransactionT]$ from $N_{j}$}
\State boolean $outcome \leftarrow$ \label{lc}
$getExclusiveLocks(T.id, T.ws)$
\NoNumber {$\wedge getSharedLocks(T.id, T.rs)$}
$\wedge validate(T.rs, T.VC)$
\If {($\neg outcome$)}
\State $releaseLocks(T.id, T.rs, T,ws)$
\State \textbf{send} \textit{Vote}$[T.id, T.VC, outcome]$ to $N_{j}$
\Else
\State $prepVC \leftarrow NLog.mostRecentVC$
\If {(${N_i} \in replicas(T.ws)$)}
\State $NodeVC[i] +  + $ \label{l9-int3}
\State $prepVC \leftarrow NodeVC$  \label{l10-int3}
\State \hspace{-0.09cm}$CommitQ.put(<T,prepVC,pending>)$
\EndIf
\State \textbf{send} \textit{Vote}$[T.id, prepVC, outcome]$ to $N_{j}$
\EndIf
\EndEvent

\Event {\textbf{receive} \textit{Decide}$[T,commitVC,outcome]$ from $N_{j}$  \textbf{atomically}} \label{l16}
\If{($outcome$)}
\State \hspace{-0.09cm}$NodeVC \leftarrow \max(NodeVC,commitVC)$
\If {(${N_i} \in replicas(T.ws)$)}
\State \hspace{-0.1cm}$CommitQ.update(<T,commitVC,ready>)$
\Else
\State $releaseSharedLocks(T.id,T.rs)$
\EndIf
\Else
\State $CommitQ.remove(T)$

\State $releaseLocks(T.id,T.ws,T.rs)$ 
\EndIf
\EndEvent \label{l29}
\Event{$\exists <T,vc,s>:<T,vc,s>  = commitQ.head \wedge s=ready$}\label{l229}
\ForAll {($k \in T.ws: {N_i} \in replicas(k)$)}
\State apply(k,val,vc)
\EndFor
\State $NLog.add( < T,vc,T.ws > )$
\State $CommitQ.remove(T)$
\State $releaseLocks(T.id,T.ws,T.rs)$
\EndEvent \label{l236}
\end{algorithmic}
\end{algorithm}
\begin{algorithm}[h]
\caption{Start Pre-commit by Transaction T in node $N_{i}$}
\label{finalize}
\begin{algorithmic}[1]
\scriptsize

\ForAll{($k \in T.ws$)} \label{ln7}
\If{(${N_i} \in replicas(k)$)}
\State $k.SQueue.insert( < T.id,vc[i],``W" > )$

\ForAll {($T^{'} \in T.PropagatedSet$)} \label{l4sp}
\State $k.SQueue.insert(<{T^{'}}.id,{T^{'}}.sid,``R">)$
\EndFor \label{l6sp}
\EndIf
\EndFor \label{ln11}
\end{algorithmic}
\end{algorithm}




Tracking only non-transitive anti-dependencies is not enough to preserve correctness. If $T$ reads the update done by $T_{w'}$ and $T_{w'}$ is still in its Pre-commit phase, then $T$ has a transitive anti-dependency with $T_{ro'}$ (i.e., $T_{ro'}\xrightarrow{\text{rw}}T_{w'}\xrightarrow{\text{wr}} T$).
SSS records the existence of transactions like $T_{ro'}$ during $T$'s execution by looking into the snapshot-queues of $T$'s read keys and logging them into a private buffer of $T$, called $T.PropagatedSet$. The propagation of anti-dependency happens during $T$'s Pre-commit phase by inserting transactions in $T.PropagatedSet$ into the snapshot-queues of all $T$'s written keys (Lines~\ref{l4sp}-\ref{l6sp} of Algorithm~\ref{finalize}).

\begin{algorithm}[h]
\caption{End Pre-commit of Transaction T in node $N_{i}$}
\label{precommit}
\begin{algorithmic}[1]
\scriptsize
\ForAll {($k \in T.ws$)} \label{le1}
\If{${N_i} \in replicas(k)$}
\State \textbf{wait until} ($\exists <{T^{'}}.id,{T^{'}}.sid,->:$ 
\NoNumber{$k.SQueue.contains(<{T^{'}}.id,{T^{'}}.sid, ->) \wedge$
\NoNumber{$ {T^{'}}.sid < T.commitVC[i]$})}
\State $k.SQueue.remove(<T.id,vc[i],``W">)$
\State \textbf{send} \textit{Ack} $[T,vc[i]]$
\textbf{to} $T.coordinator$
\EndIf
\EndFor \label{le7}
\end{algorithmic}
\end{algorithm}

\underline{\textit{External Commit}}.
Transaction $T$ remains in its Pre-commit phase until there is no read-only transaction with lesser insertion-snapshot in the snapshot-queues of $T$'s written keys. After that, $T$ is removed from these snapshot-queues and an \texttt{Ack} message to the transaction 2PC coordinator is sent (Lines~\ref{le1}-\ref{le7} of Algorithm~\ref{precommit}).

The coordinator can inform its client after receiving \texttt{Ack} from all 2PC participants.
At this stage, update transaction's external schedule is established, therefore we say that SSS commits the update transaction \textit{externally}.

\subsection{Execution of Read-Only Transactions}

In its first read operation (Algorithm~\ref{read} Lines~\ref{firsta}-\ref{firstb}), a read-only transaction $T$ on $N_i$ assigns \texttt{NLog.mostRecentVC} to its vector clock ($T.VC$). This way, $T$ will be able to see the latest updated versions committed on $N_i$.
Read operations are implemented by contacting all nodes that replicate the requested key and waiting for the fastest to answer.

When a read request of $T$ returns from node $N_j$, $T$ sets $T.hasRead[j]$ to \texttt{true}. With that, we set the visibility upper bound for $T$ from $N_{j}$ (i.e., $T.VC[j]$). Hence, subsequent read operations by $T$ contacting a node $N_{k}$ should only consider versions with a vector clock $vc_{k}$ such that $vc_{k}[j]<T.VC[j]$.


After a read operation returns, the transaction vector clock is updated by applying an entry-wise maximum operation between the current $T.VC$ and the vector clock associated with the read version (i.e., $VC^{*}$) from $N_{j}$.
Finally, the read value is added to $T.rs$ and returned.

Algorithm~\ref{visibility} shows SSS rules to select the version to be returned upon a read operation that contacts node $N_i$.


The first time $N_{i}$ receives a read from $T$, this request should wait until the value of $N_i$'s \texttt{NLog.mostRecentVC[i]} is equal to $T.VC[i]$
(Line~\ref{line4} Algorithm~\ref{visibility}).
This means that all transactions that are already included in the current visibility bound of $T.VC[i]$ must perform their internal commit before $T$'s read request can be handled.

After that, a correct version of the requested key should be selected for reading. This process starts by identifying the set of versions that are within the visibility bound of $T$, called $VisibleSet$. This means that, given a version $v$ with commit vector clock $vc$, $v$ is visible by $T$ if, for each entry $k$ such that $T.hasRead[k]=true$, we have that $vc[k]\leq T.VC[k]$ (Algorithm~\ref{visibility} Line~\ref{vis-line5}).


\begin{algorithm}[h]
\scriptsize
\caption {Read Operation by Transaction T in node $N_{i}$}
\label{read}
\begin{algorithmic}[1]
\scriptsize
\Event  {Value Read(Transaction T, Key k)}
\If{($\exists  < k,val >  \in T.ws$)}
\State \Return{val}
\EndIf
\If {(is first read of T)} \label{firsta}
\State $T.VC \leftarrow NLog.mostRecentVC$ 
\EndIf \label{firstb}
\State ${\rm{target}} \leftarrow {\rm{\{ replicas(k)}}\} $
\State\textbf{send} \textit{READREQUEST}$[k, T.VC,T.hasRead,T.isUpdate]$ 
\NoNumber {to \textbf{all} ${N_j} \in target$}
\State \textbf{wait Receive}  \textit{READRETURN} $[val, VC^{*}, PropagatedSet]$ 
\Statex \hspace{0.4cm} from {${N_h} \in target$}
\State $T.hasRead[h] \leftarrow true$
\State $T.VC \leftarrow \max (T.VC,V{C^{*}})$ \label{l12-read}
\State $T.rs \leftarrow T.rs \cup \{ <k,val>\} $
\State $T.PropagatedSet \leftarrow T.PropagatedSet \cup PropagatedSet$
\State \Return {val}
\EndEvent
\end{algorithmic}
\end{algorithm}
\begin{algorithm}[h]
\caption{Version Selection Logic in node $N_{i}$}
\label{visibility}
\begin{algorithmic}[1]
\scriptsize
\Event{Receive \textit{READREQUEST}$[T, k, T.VC, hasRead, isUpdate]$ from $N_{j}$}   
\State $PropagatedSet \leftarrow \phi$
\If {($\neg isUpdate$)}
\If {($\neg {\rm{hasRead[i]}}$)}
\State \textbf{wait until} \label{line4} ${\rm{NLog}}{\rm{.mostRecentVC[i]}} \ge$ ${\rm{T.VC[i]}}$ 
\State $VisibleSet \leftarrow \{ vc: vc   \in {\rm{NLog}} \wedge$ \label{vis-line5}
 
\Statex \hspace{1.2cm} $\forall w({\rm{hasRead[w]= true}}$ $\Rightarrow {\rm{vc[w]}}$  $\le {\rm{T.VC[w]}})\} $

\State $ExcludedSet\leftarrow\{ T^{'}:<T^{'}.id,T^{'}.cid,``W"> \in$  \label{ver-l6}
\Statex \hspace{1.2cm}$ k.SQueue  \Rightarrow vc[i] > T.VC[i]) \}$
\State $VisibleSet \leftarrow VisibleSet \backslash ExcludedSet$

\State$maxVC \leftarrow vc:\forall w,vc[w] = \max \{ v[w]:v$ $\in VisibleSet\} $
\State $k.SQueue.insert( <T.id,maxVC[i],``R"> )$

\State $ver \leftarrow k.last$
\While{$(\exists w:hasRead[w] = true \wedge ver.vc[w] >$ 
\Statex  \indent \indent \indent  $maxVC[w]\vee \exists vc \in ExcludedSet:ver.vc = vc $
\Statex  \indent \indent \indent $\wedge vc[i] > maxVC[i])$}
\State $ver \leftarrow ver.prev$
\EndWhile 
\Else  \label{vis-line15}
\State $maxVC \leftarrow T.VC$
\State $k.SQueue.insert(< T.id,maxVC[i],``R">)$
\State $ver \leftarrow k.last$
\While{($\exists w:({\rm{hasRead[w] = }}true \wedge ver.vc{\rm{[w]}} > maxVC[w])$)}
\State $ver \leftarrow ver.prev$
\EndWhile
\EndIf \label{line20}
\Else 
\State $maxVC \leftarrow NLog.mostRecentVC$ \label{vis24}
\State $PropagatedSet$=$\{T^{'}:<T^{'}.id,readSid,``R">\in k.SQueue\}$
\State $ver \leftarrow k.last$
\EndIf                              \label{vis26}      
\State \textbf{Send} \textit{READRETURN}[ver.val, maxVC,PropagatedSet ] \textbf{to} $N_{j}$
\EndEvent
\end{algorithmic}
\end{algorithm}

It is possible that transactions associated with some of these vector clocks are still in their Pre-Commit phase, meaning they exist in the snapshot-queues of $T$'s requested key. If so, they should be excluded from $VisibleSet$ in case their insertion-snapshot is higher than $T.VC[i]$.
The last step is needed
to serialize read-only transactions with anti-dependency relations before conflicting update transactions.

This condition is particularly important to prevent a well-known anomaly, firstly observed by Adya in~\cite{adya}, in which read-only transactions executing on different nodes can observe two non-conflicting update transactions in different serialization order~\cite{gmu12}. 
Consider a distributed system where nodes do not have access to a single point of synchronization (or an ordering component), concurrent non-conflicting transactions executing on different nodes cannot be aware of each other's execution. Because of that, different read-only transactions might order these non-conflicting transactions in a different way, therefore breaking the client's perceived order. SSS prevents that by serializing both these read-only transactions before those update transactions.

At this stage, if multiple versions are still included in $VisibleSet$, the version with the maximum $VC[i]$ should be selected to ensure external consistency.



Once the version to be returned is selected, $T$ is inserted in the snapshot-queue of the read key using
$maxVC[i]$ as insertion-snapshot (Line 16 of Algorithm~\ref{visibility}). Finally, when the read response is received, the maximum per entry between $maxVC$ (i.e., $VC^{*}$ in Algorithm~\ref{read}) and the $T.VC$  is computed along with the result of the read operation.

When a read-only transaction $T$ commits, it immediately replies to its client. After that, it sends a message to the nodes storing only the read keys in order to notify its completion. We name this message \texttt{Remove}.
Upon receiving \texttt{Remove}, the read-only transaction is deleted from all the snapshot-queues associated with the read keys.
Deleting a read-only transaction from a snapshot-queue enables conflicting update transactions to be externally committed and their responses to be released to their clients.

Because of transitive anti-dependency relations, a node might need to forward the \texttt{Remove} message to other nodes as follows. Let us assume $T$ has an anti-dependency with a transaction $T_w$, and another transaction $T_{w'}$ reads from $T_w$. Because anti-dependency relations are propagated along the chain of conflicting transactions, $T$ exists in the snapshot-queues of $T_{w'}$'s written keys. Therefore, upon \texttt{Remove} of $T$, the node executing $T_w$ is responsible to forward the \texttt{Remove} message to the node where $T_{w'}$ executes for updating the affected snapshot-queues.




When a read operation is handled by a node that already responded to a previous read operation from the same transaction, 
the latest version according to $maxVC$
is returned,
and $T$ can be inserted into the snapshot-queue with its corresponding identifier and $maxVC[i]$ as insertion-snapshot.

\subsection{Examples} \label{example}
\begin{figure}[h]
\centering
    \includegraphics[scale=0.3]{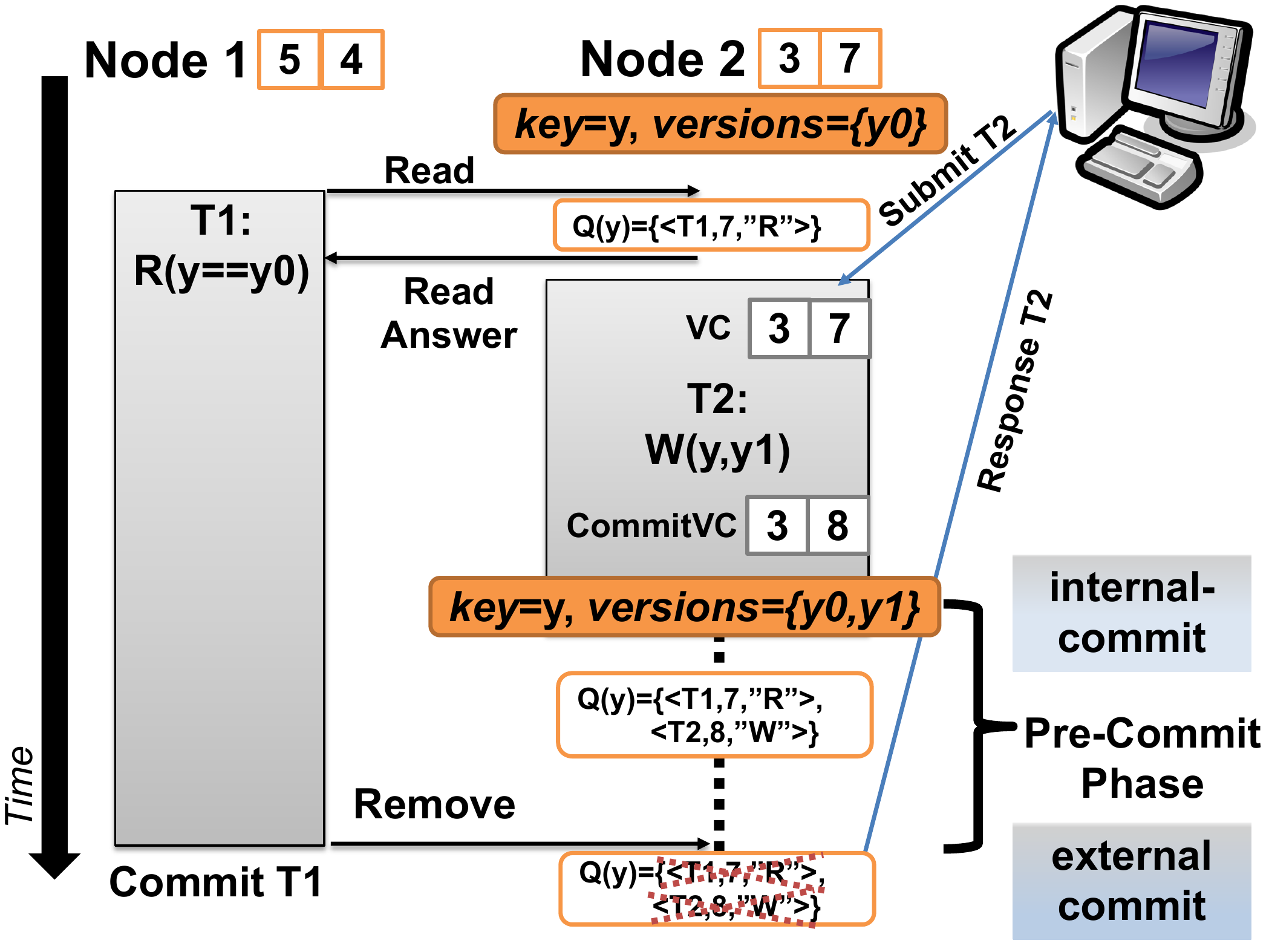}
    \caption{SSS execution in the presence of an anti-dependency. Orange boxes show the content of the data store. Gray boxes show transaction execution. Dashed line represents the waiting time for $T2$. The red crossed entries of Q(y) represent their elimination upon Remove.}
\label{fig1}
\end{figure}

\textbf{External Consistency \& Anti-dependency}.
Figure~\ref{fig1} shows an example of how SSS serializes an update transaction $T_{1}$ in the presence of a concurrent read-only transaction $T_{2}$. Two nodes are deployed, $N_{1}$ and $N_{2}$, and no replication is used for simplicity. $T_{1}$ executes on $N_{1}$ and $T_{2}$ on $N_{2}$. Key $y$ is stored in $N_{2}$'s repository. The \texttt{NLog.mostRecentVC} for Node 1 is [5,4] and for Node 2 is [3,7].

$T_{1}$ performs a read operation on key $y$ by sending a remote read request to $N_{2}$. At this point, $T_{1}$ is inserted in the snapshot-queue of $y$ ($Q(y$)) with 7 as insertion-snapshot. This value is the second entry of $N_{2}$'s \texttt{NLog.mostRecentVC}.
Then the update transaction $T_{2}$ begins with vector clock [3,7], buffers its write on key $y$ in its write-set, and performs its internal commit by making the new version of $y$ available, and by inserting the produced commit vector clock (i.e., $T2.commitVC$=[3,8]) in $N_{2}$'s $NLog$. As a consequence of that, $NLog.mostRecentVC$
is equal to $T2.commitVC$.

Now $T_2$ is evaluated to decide whether it should be inserted into $Q(y)$. The insertion-snapshot of $T_2$ is equal to 8, which is higher than $T_{1}$'s insertion-snapshot in $Q(y)$. For this reason, $T_2$ is inserted in $Q(y)$ and its Pre-commit phase starts.


At this stage, $T_{2}$ is still not externally visible. Hence $T_{2}$ remains in its Pre-Commit phase until $T_{1}$ is removed from $Q(y)$, which happens when $T_1$ commits and sends the \texttt{Remove} message to $N_2$. After that, $T_{2}$'s client is informed about $T_{2}$'s completion. Delaying the external commit of $T_{2}$ shows clients a sequence of transactions completion that matches their serialization order.



\begin{figure}[h]
  \centering
\includegraphics[scale=0.25]{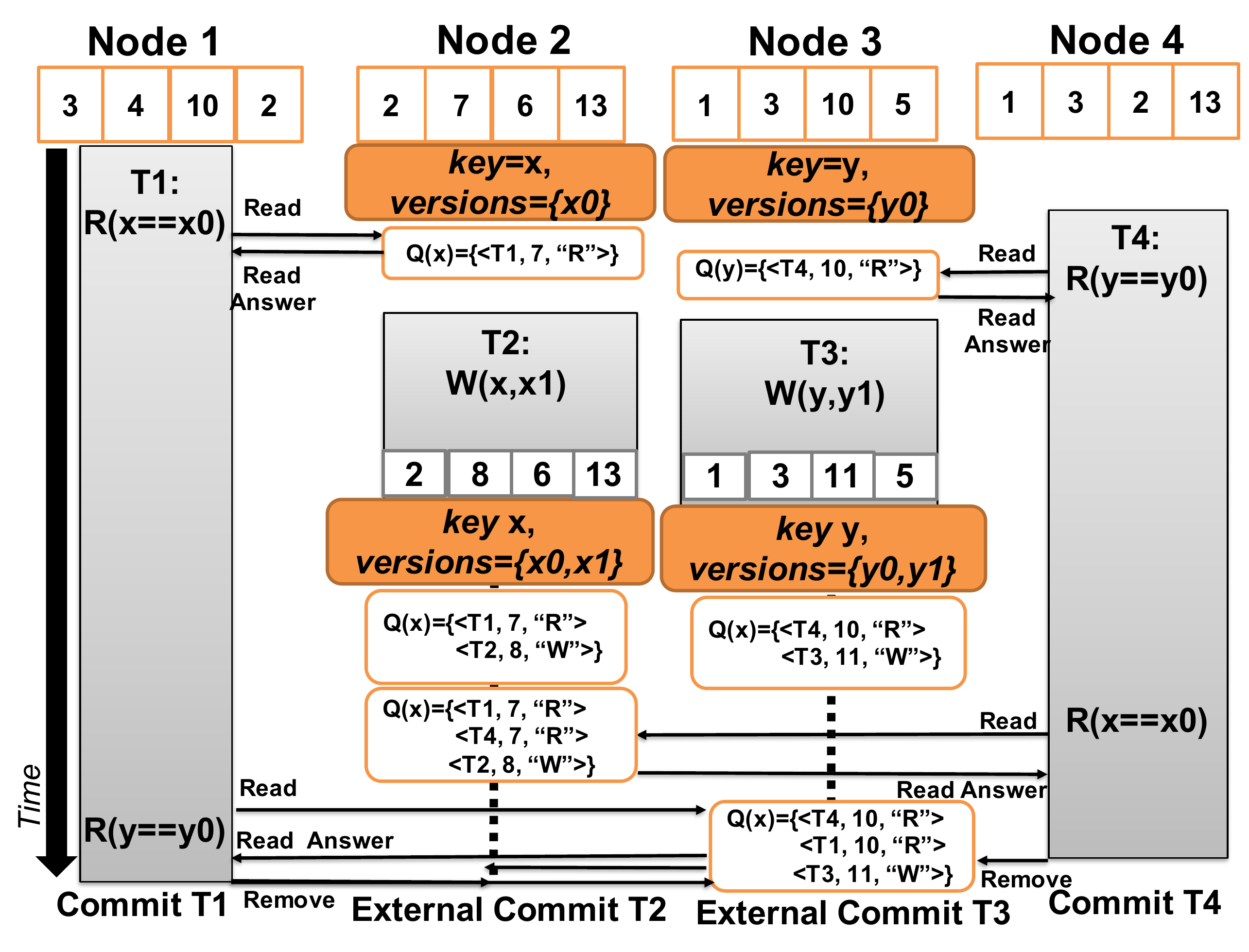}
    \caption{Handling read-only transactions along with non-conflicting update transactions. We omitted snapshot-queue entries elimination upon Remove to improve readability.}
\label{pl3u-fixed}
\end{figure}

\textbf{External Consistency \& Non-conflicting transactions}.
Figure~\ref{pl3u-fixed} shows how SSS builds the external schedule in the presence of read-only transactions and non-conflicting update transactions.
There are four nodes, $N_1$, $N_2$, $N_3$, $N_4$, and four concurrent  transactions, $T_1$, $T_2$, $T_3$, $T_4$, each executes on the respective node. By assumption, $T_2$ and $T_3$ are non-conflicting update transactions, while $T_1$ and $T_4$ are read-only.

SSS ensures that $T_1$ and $T_4$ do not serialize $T_2$ and $T_3$ in different orders and they return to their clients in the same way they are serialized by relying on snapshot-queuing. $T_1$ is inserted into $Q(x)$ with insertion-snapshot equals to 7. Concurrently, $T_4$ is added to the snapshot-queue of $y$ with insertion-snapshot equals to 10.
The next read operation by $T_1$ on $y$ has two versions evaluated to be returned: $y0$ and $y1$. Although $y1$ is the most recent, since $T_4$ returned $y0$ previously (in fact $T_4$ is in $Q(y)$), $y1$ is excluded and $y0$ is returned. Similar arguments apply to $T_4$'s read operation on $x$.
The established external schedule serializes $T_1$ and $T_4$ before both $T_2$ and $T_3$.

\subsection{Additional Considerations of SSS}

\underline{\textit{Garbage Collection}}. A positive side effect of the \texttt{Remove} message is the implicit garbage collection of entries in the snapshot-queues. In fact, SSS removes any entry representing transactions waiting for a read-only transaction to finish upon receiving \texttt{Remove}, which cleans up the snapshot-queues.

\underline{\textit{Starvation}}. Another important aspect of SSS is the chance to slow down update transactions, possibly forever, due to an infinite chain of conflicting read-only transactions issued concurrently. We handle this corner case by applying admission control to read operations of read-only transactions in case they access a key written by a transaction that is in a snapshot-queue for a pre-determined time. In practice, if such a case happens, we apply an artificial delay to the read operation (exponential back-off) to give additional time to update transaction to be removed from the snapshot-queue. In the experiments we never experienced starvation scenarios, even with long read-only transactions.


\underline{\textit{Deadlock-Freedom}}. SSS uses timeout to prevent deadlock during the commit phase's lock acquisition. Also, the waiting condition applied to update transactions cannot generate deadlock. This is because read-only transactions never wait for each other, and there is no condition in the protocol where an update transaction blocks a read-only transaction. The only wait condition occurs when read-only transactions force update transactions to hold their client response due to snapshot-queuing. As a result, no circular dependency can be formed, thus SSS cannot encounter deadlock.

\underline{\textit{Fault-Tolerance}}. SSS deploys a protocol that tolerates failures in the system using replication. In the presented version of the SSS protocol, we did not include either logging of messages to recover update transactions' 2PC upon faults, or a consensus-based approach (e.g., Paxos-Commit~\cite{2pc}) to distribute and order 2PC messages. Solutions to make 2PC recoverable are well-studied. To focus on the performance implications of the distributed concurrency control of SSS and all its competitors, operations to recover upon a crash of a node involved in a 2PC have been disabled. This decision has no correctness implication.


\section{Correctness}
\label{sec:correctness}

Our target is proving that every history $H$ executed by SSS, which includes committed update transactions and read-only transactions (committed or not), is external consistent.

We adopt the classical definition of history~\cite{adya}.
For understanding correctness, it is sufficient to know that a history is external consistent if the transactions in the history return the same values and leave the data store in the same state as they were executed in a sequential order (one after the other), and that order does not contradict the order in which transactions return to their clients.

We decompose SSS's correctness in three statements, each highlighting a property guaranteed by SSS.
Each statement claims that a specific history $H'$, which is derived from $H$, is external consistent. In order to prove that, we rely on the characteristics of the Direct Serialization Graph (DSG)~\cite{adya} which is derived from $H'$. Note that DSG also includes order relations between transactions' completion.

Every transaction in $H'$ is a node of the DSG graph, and every dependency of a transaction $T_j$ on a transaction $T_i$ in $H'$ is an edge from $T_i$ to $T_j$ in the graph. The concept of dependency is the one that is widely adopted in the literature: \textit{i)} $T_j$ read-depends on $T_i$ if a read of $T_j$ returns a value written by $T_i$, \textit{ii)} $T_j$ write-depends on $T_i$ if a write of $T_j$ overwrites a value written by $T_i$; \textit{iii)} $T_j$ anti-depends on $T_i$ if a write of $T_j$ overwrites a value previously read by $T_i$. We also map the completion order relations to edges in the graph: if $T_i$ commits externally before $T_j$ does, then the graph has an edge from $T_i$ to $T_j$. A history $H'$ is external consistent iff the DSG does not have any cycle~\cite{serializability,adya}. 

In our proofs we use the binary relation $\leq$ to define an ordering on pair of vector clocks $v_1$ and $v_2$ as follows: $v_1 \leq v_2$ if $\forall i$, $v_1[i]\leq v_2[i]$. Furthermore, if there also exists at least one index $j$ such that $ v_1[j]<v_2[j]$, then $v_1<v_2$ holds. 

\textit{\underline{Statement 1}. For each history $H$ executed by SSS, the history $H'$, which is derived from $H$ by only including committed update transactions in $H$, is external consistent.}

In the proof we show that if there is an edge from transaction $T_i$ to transaction $T_j$ in DSG, then $T_i.commitVC$ $<$ $T_j.commitVC$. This statement implies that transactions modify the state of the data store as they were executed in a specific sequential order, which does not contradict the transaction external commit order. Because no read-only transactions is included in $H'$, the internal commit is equivalent to the external commit (i.e., no transaction is delayed). The formal proof is included in the technical report~\cite{tech-rep}.


\textit{\underline{Statement 2}. For each history $H$ executed by SSS, the history $H'$, which is derived from $H$ by only including committed update transactions and one read-only transaction in $H$, is external consistent.}

The proof shows that read-only transactions always observe a consistent state by showing that in both the case of a direct dependency or anti-dependency, the vector clock of the read-only transactions is comparable with the vector clocks of conflicting update transactions.
This statement implies that read operations of a read-only transaction always return values from a state of the data store as the transaction was executed atomically in a point in time that is not concurrent with any update transaction. The formal proof is included in the technical report~\cite{tech-rep}.

\textit{\underline{Statement 3}. For each history $H$ executed by SSS, the history $H'$, which is derived from $H$ by only including committed update transactions and two or more read-only transactions in $H$, is external consistent.}

Since Statement 2 holds, SSS guarantees that each read-only transaction appears as it were executed atomically in a point in time that is not concurrent with any update. Furthermore, since Statement 1 holds, the read operations of that transaction return values of a state that is the result of a sequence of committed update transactions. Therefore, Statement 3 implies that, given such a sequence $S1$ for a read-only transaction $T_{r1}$, and $S2$ for a read-only transaction $T_{r2}$, either $S1$ is a prefix of $S2$, or $S2$ is a prefix of $S1$.
In practice, this means that all read-only transactions have a coherent view of all transactions executed on the system. The formal proof is included in the technical report~\cite{tech-rep}.

\begin{figure*}[!htb]

\begin{minipage}{.64\textwidth}
\centering
\includegraphics[width=.9\textwidth]{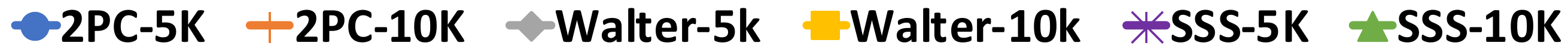}\\
\hspace{-10pt}
\subfigure[20\%]{
    \includegraphics[height=72pt,keepaspectratio]{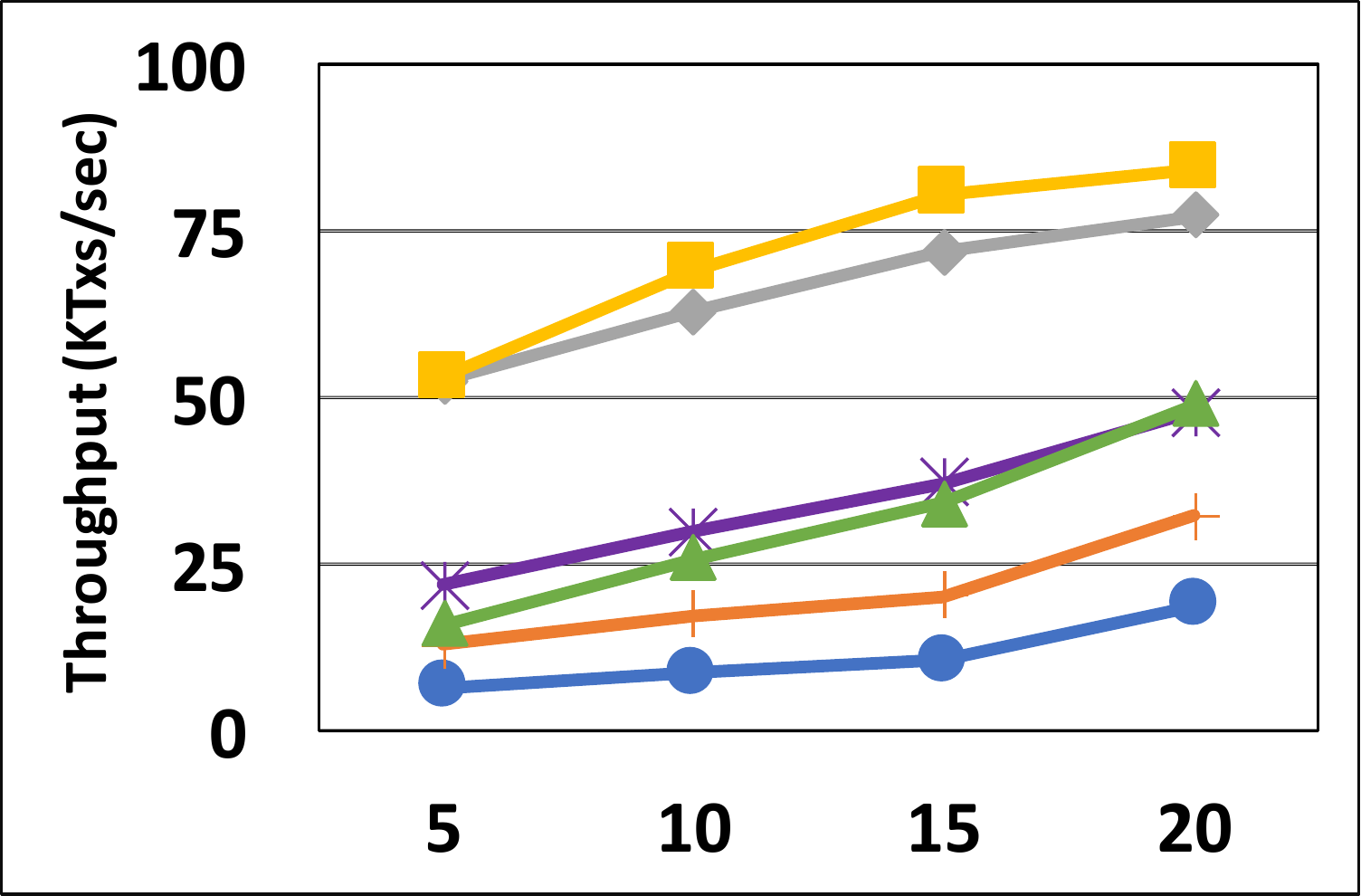}
    \label{fig:repl-thr-nolocal-a}
}\hspace{-10pt}
\subfigure[50\%]{
    \includegraphics[height=72pt,keepaspectratio]{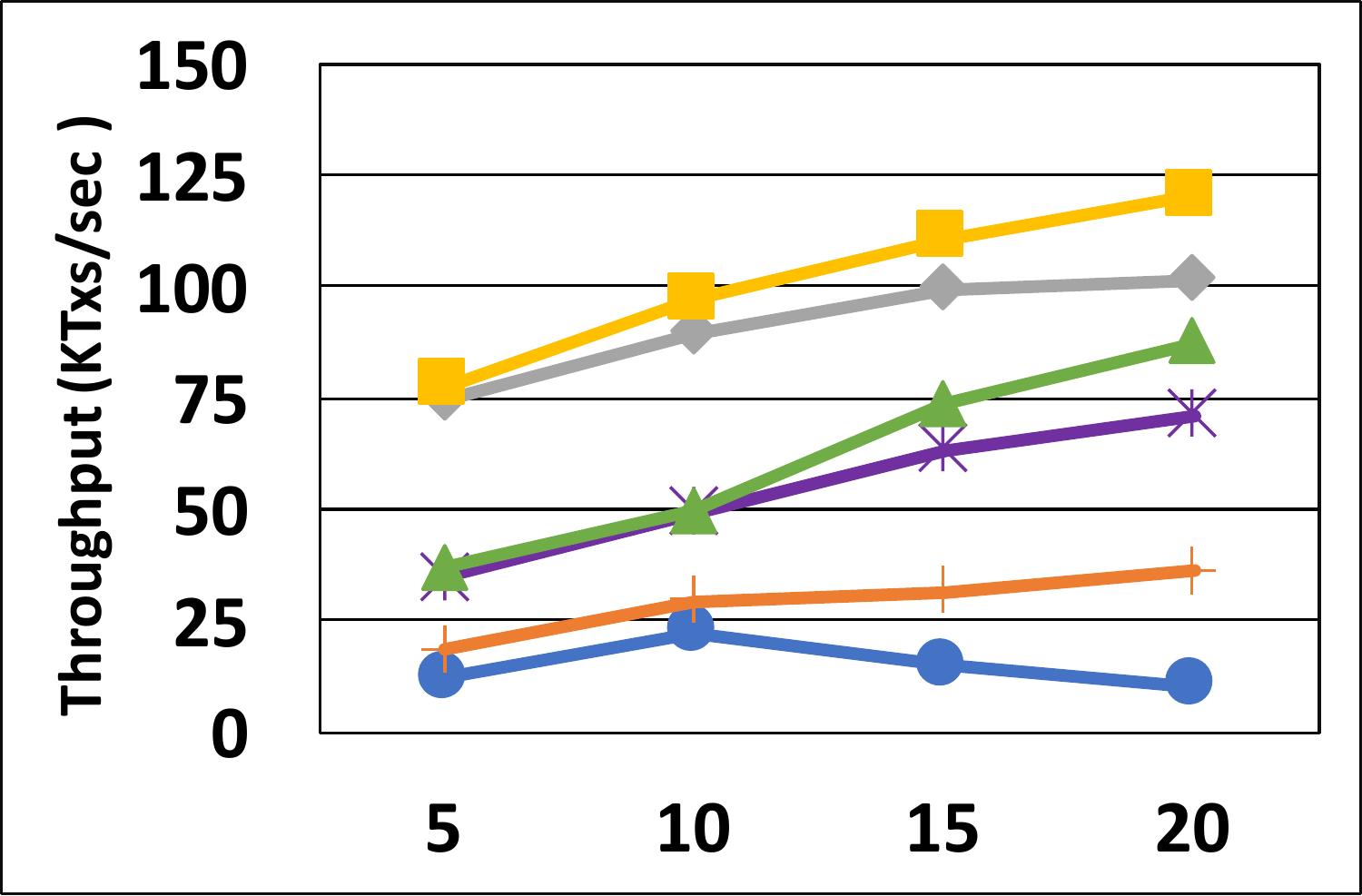}
    \label{fig:repl-thr-nolocal-b}
}\hspace{-10pt}
\subfigure[80\%]{
    \includegraphics[height=72pt,keepaspectratio]{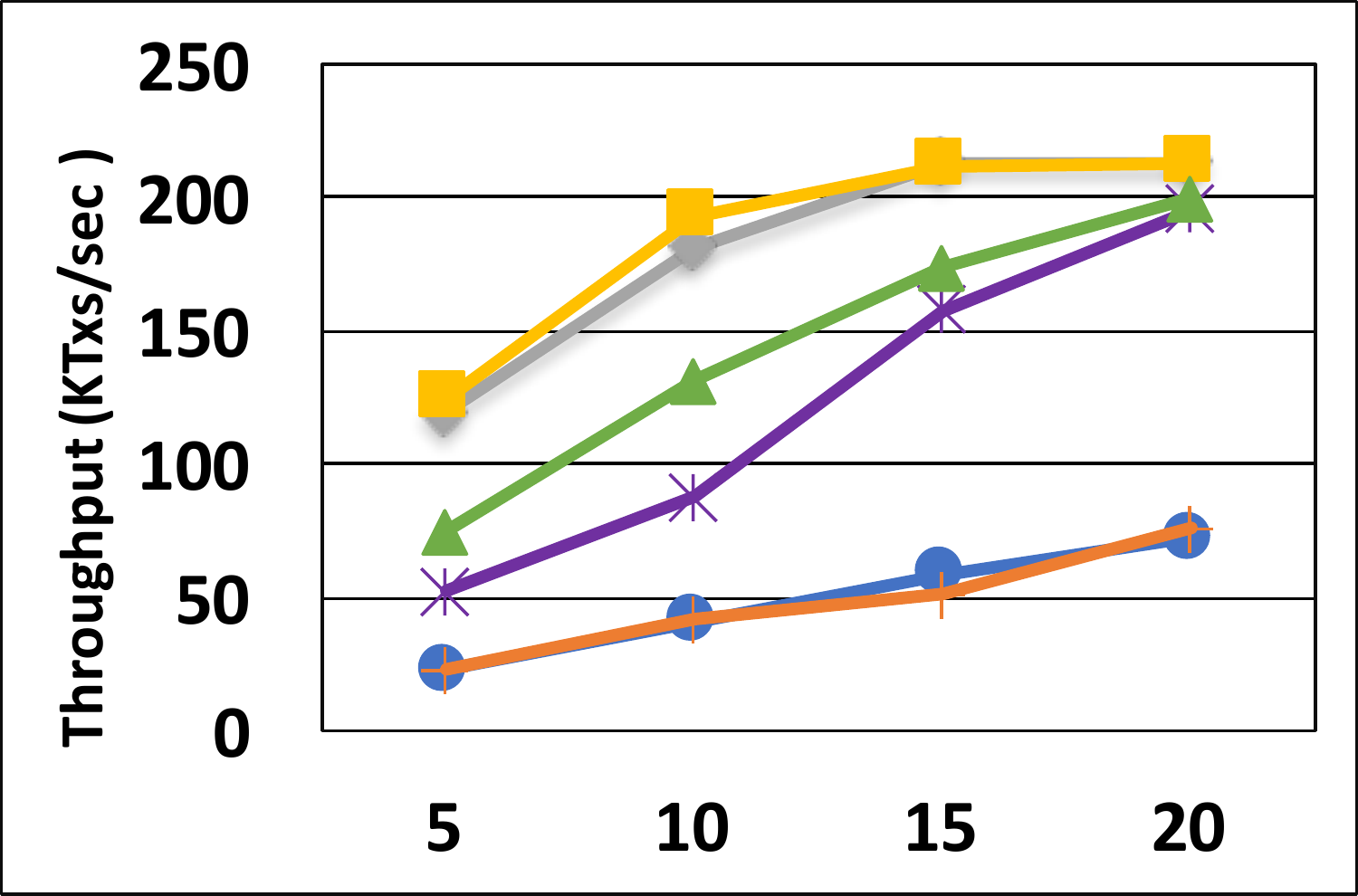}
    \label{fig:repl-thr-nolocal-c}
}
\vspace{-4pt}
\caption{Throughput varying \% of read-only transactions. Number of nodes in X-axes.}
\label{fig:repl-thr-nolocal}
\end{minipage}\hfill
\begin{minipage}{0.36\textwidth}
\centering
\subfigure[Maximum attainable throughput. Number of nodes in X-axis.]{
    \includegraphics[height=70pt,keepaspectratio]{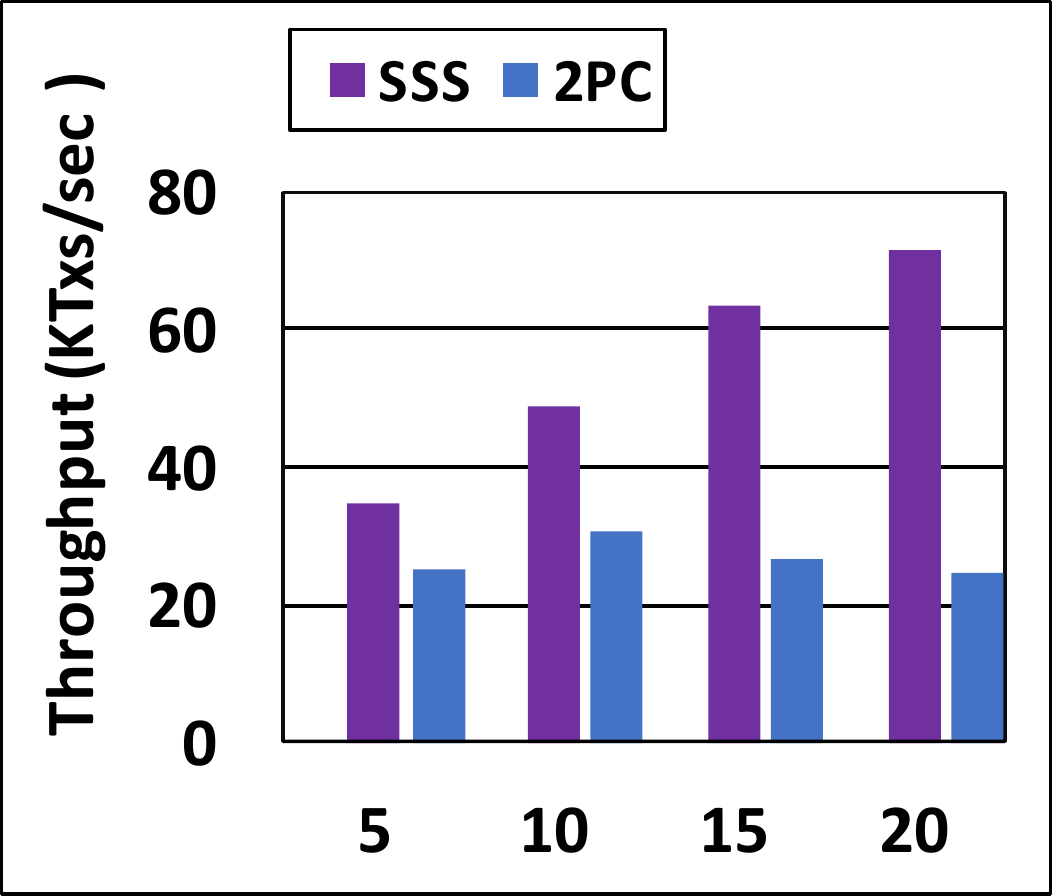}
    \label{fig:max-thr}
}\hspace{2pt}
\subfigure[External Commit latency. Clients per node in X-axis.]{
    \includegraphics[height=70pt,keepaspectratio]{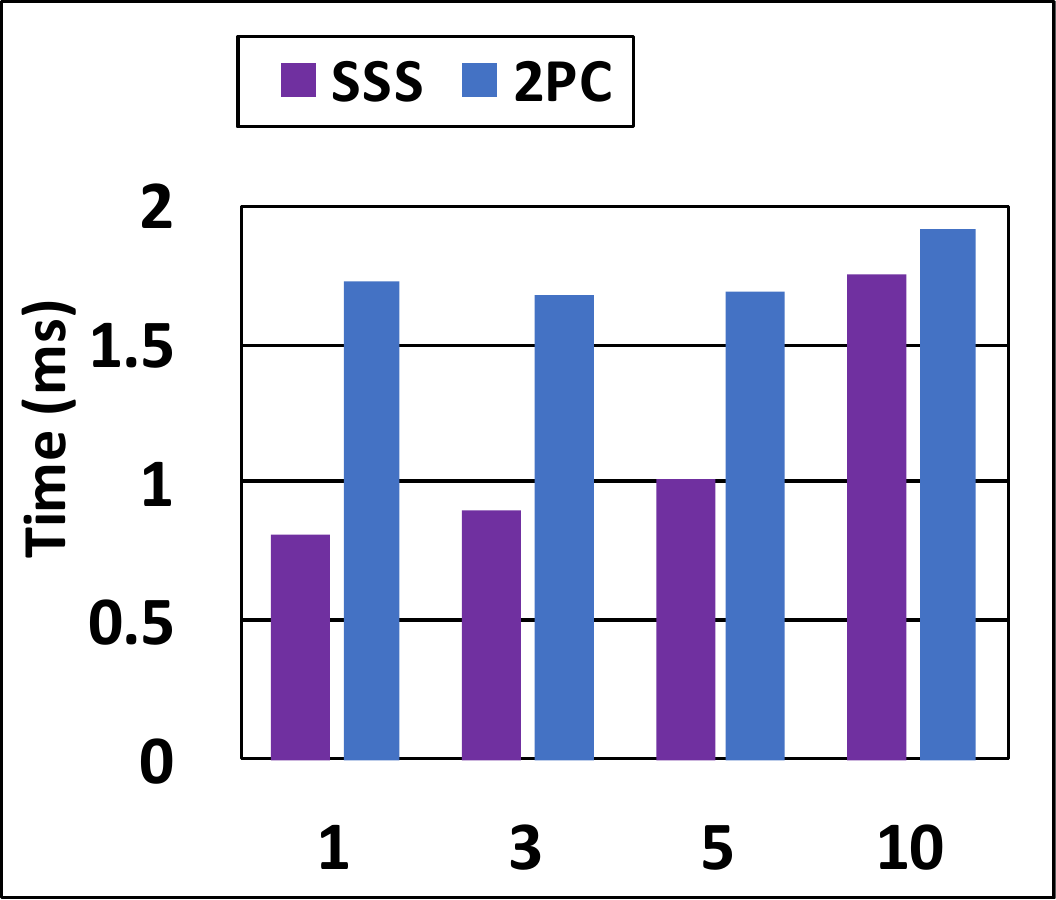}
    \label{fig:latency}
}
\vspace{-10pt}
\caption{Performance of SSS against 2PC-baseline using 5k objects and 50\% read-only transactions.}
\label{myass}
\end{minipage}
\end{figure*}

\section{Evaluation} \label{sec:eval}

We implemented SSS in Java from the ground up and performed a comprehensive evaluation study. In the software architecture of SSS there is an optimized network component where multiple network queues, each for a different message type, are deployed. This way, we can assign priorities to different messages and avoid protocol slow down in some critical steps due to network congestion caused by lower priority messages (e.g., the \texttt{Remove} message has a very high priority because it enables external commits). Another important implementation aspect is related to snapshot-queues. Each snapshot-queue is divided into two: one for read-only transactions and one for update transactions. This way, when the percentage of read-only transactions is higher than update transactions, a read operation should traverse few entries in order to establish its visible-set.

We compare SSS against the following competitors: 2PC-baseline (shortly 2PC in the plots), ROCOCO~\cite{rococo}, and Walter~\cite{walter}. All these competitors offer transactional semantics over key-value APIs. With 2PC-baseline we mean the following implementation: all transactions execute as SSS's update transactions; read-only transactions validate their execution, therefore they can abort; and no multi-version data repository is deployed. As SSS, 2PC-baseline  guarantees external consistency.

ROCOCO is an external consistent  two-round protocol where transactions are divided into pieces and dependencies are collected to establish the execution order. ROCOCO classifies pieces of update transactions into immediate and deferrable. The latter are more efficient because they can be reordered. Read-only transactions can be aborted, and they are implemented by waiting for conflicting transactions to complete. Our benchmark is configured in a way all pieces are deferrable. ROCOCO uses preferred nodes to process transactions and consensus to implement replication. Such a scheme is different from SSS where multiple nodes are involved in the transaction commit process.
To address this discrepancy, in the experiments where we compare SSS and ROCOCO, we disable replication for a fair comparison.
The third competitor is Walter, which provides PSI a weaker isolation level than SSS. Walter has been included because it synchronizes nodes using vector clocks, as done by SSS.

All competitors have been re-implemented using the same software infrastructure of SSS because we want to provide all competitors with the same underlying code structure and optimization (e.g., optimized network). For fairness, we made sure that the performance obtained by our re-implementation of competitors matches the  trends reported in~\cite{walter} and~\cite{rococo}, when similar configurations were used.

In our evaluation we use YCSB~\cite{DBLP:conf/cloud/CooperSTRS10} benchmark ported to key-value store. We configure the benchmark to explore multiple scenarios. We have two transaction profiles: update, where two keys are read and written, and read-only transactions, where two or more keys are accessed.
In all the experiments we co-locate application clients with processing nodes, therefore increasing the number of nodes in the system also increases the amount of issued requests. There are 10 application threads (i.e., clients) per node injecting transactions in the system in a closed-loop (i.e., a client issues a new request only when the previous one has returned).
All the showed results are the average of 5 trials.

We selected two configurations for the total number of shared keys: 5k and 10k. With the former, the observed average transaction abort rate is in the range of 6\% to 28\% moving from 5 nodes to 20 nodes when 20\% read-only transactions are deployed. In the latter, the abort rate was from 4\% to 14\%. Unless otherwise stated, transactions select accessed objects randomly with uniform distribution.

As test-bed, we used CloudLab~\cite{ricci2014introducing}, a cloud infrastructure available to researchers. We selected 20 nodes of type c6320 available in the Clemson cluster~\cite{clamson}. This type is a physical machine with 28 Intel Haswell CPU-cores and  256GB of RAM. Nodes are interconnected using 40Gb/s Infiniband HPC cards.
In such a cluster, a network message is delivered in around 20 microseconds (without network saturation), therefore we set timeout on lock acquisition to 1ms.



In Figure~\ref{fig:repl-thr-nolocal} we compare the throughput of SSS against 2PC-baseline and Walter in the case where each object is replicated in two nodes of the system. We also varied the percentage of read-only transactions in the range of 20\%, 50\%, and 80\%.
As expected, Walter is the leading competitor in all the scenarios because its consistency guarantee is much weaker than external consistency; however, the gap between SSS and Walter reduces from 2$\times$ to 1.1$\times$ when read-only transactions become predominant (moving from Figure~\ref{fig:repl-thr-nolocal-a} to~\ref{fig:repl-thr-nolocal-c}). This is reasonable because in Walter, update transactions do not have the same impact in read-only transactions' performance as in SSS due to the presence of the snapshot-queues. Therefore, when the percentage of update transactions reduces, SSS reduces the gap. Considering the significant correctness level between PSI (in Walter) and external consistency, we consider the results of the comparison between SSS and Walter remarkable.

Performance of 2PC-baseline is competitive when compared with SSS only at the case of 20\% read-only. In the other cases, although SSS requires a more complex logic to execute its read operations, the capability of being abort-free allows SSS to outperform 2PC-baseline by as much as 7$\times$ with 50\% read-only and 20 nodes. 2PC-baseline's performance in both the tested contention levels become similar at the 80\% read-only case because, although lock-based, read-only transaction's validation will likely succeed since few update transactions execute in the system.

Figure~\ref{fig:repl-thr-nolocal} also shows the scalability of all competitors. 2PC-baseline suffers from higher abort rate than others, which hampers its scalability. This is because its read-only transactions are not abort-free. The scalability trend of SSS and Walter is similar, although Walter stops scaling at 15 nodes using 80\% of read-only transactions while SSS proceeds. This is mostly related with network congestion, which is reached by Walter earlier than SSS since Walter's transaction processing time is lower than SSS, thus messages are sent with a higher rate.

In Figure~\ref{myass} we compare 2PC-baseline and SSS in terms of maximum attainable throughput and transaction latency. Figure~\ref{fig:max-thr} shows 2PC-baseline and SSS configured in a way they can reach their maximum throughput with 50\% read-only workload and 5k objects, meaning the number of clients per nodes differs per reported datapoint. Performance trends are similar to those in Figure~\ref{fig:repl-thr-nolocal-b}, but 2PC-baseline here is faster than before. This is related with the CPU utilization of the nodes' test-bed. In fact, 2PC-baseline requires less threads to execute, meaning it leaves more unused CPU-cores than SSS, and those CPU-cores can be leveraged to host more clients.


\begin{figure}[h]
\centering
\includegraphics[width=.2\textwidth]{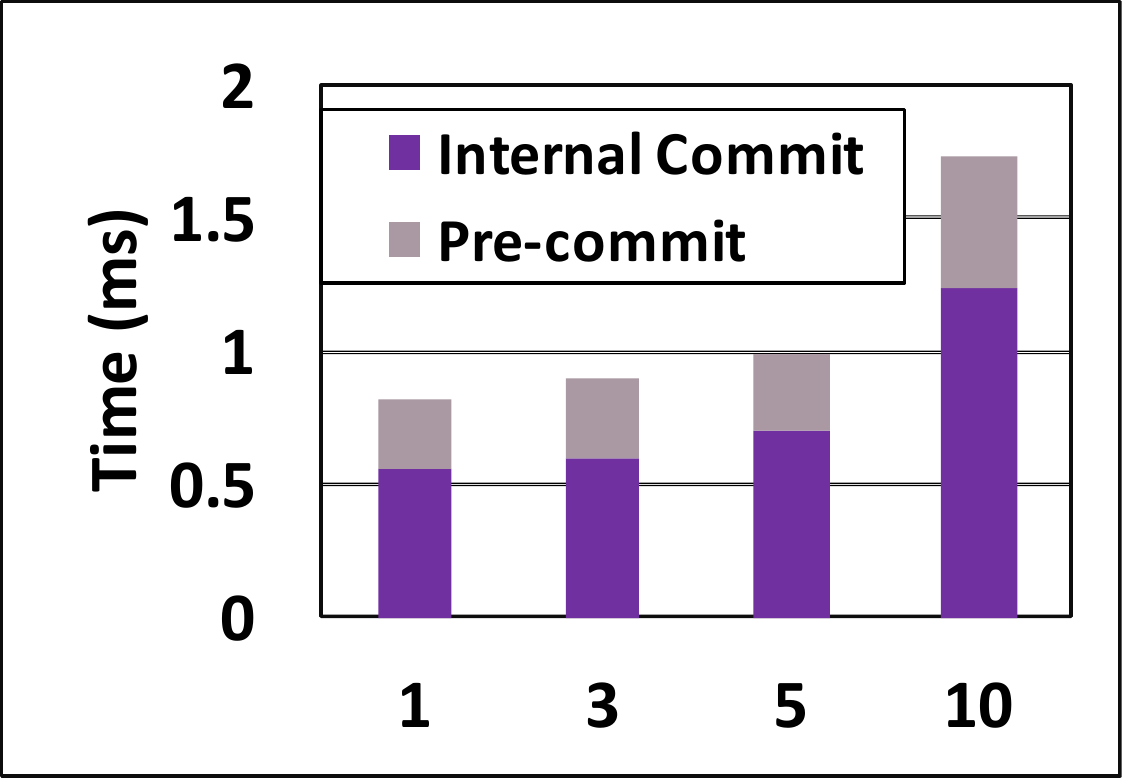}
\caption{Breakdown of SSS transaction latency.}
\label{fig:breakdown}
\end{figure}

The second plot (Figure~\ref{fig:latency}) shows transaction latency from its begin to its external commit when 20 nodes, 50\% read-only transactions, and 5k objects are deployed.
In the experiments we varied the number of clients per node from 1 to 10. When the system is far from reaching saturation (i.e., from 1 to 5 clients), SSS's latency does not vary, and it is on average 2$\times$ lower than 2PC-baseline's latency. At 10 clients, SSS's latency is still lower than 2PC-baseline but by a lesser percentage. This confirms one of our claim about SSS capability of retaining high-throughput even when update transactions are held in snapshot-queues. In fact, Figure~\ref{fig:repl-thr-nolocal-b} shows the throughput measurement in the same configuration: SSS is almost 7$\times$ faster than 2PC-baseline.

Figure~\ref{fig:breakdown} shows the relation between the internal commit latency and the external commit latency of SSS update transactions. The configuration is the one in Figure~\ref{fig:latency}. Each bar represents the latency between a transaction begin and its external commit. The internal red bar shows the time interval between the transaction's insertion in a snapshot-queue and its removal (i.e., from internal to external commit). This latter time is on average 30\% of the total transaction latency.

\begin{figure}[h]
\centering
\subfigure[20\%.]{
    \includegraphics[width=.2\textwidth]{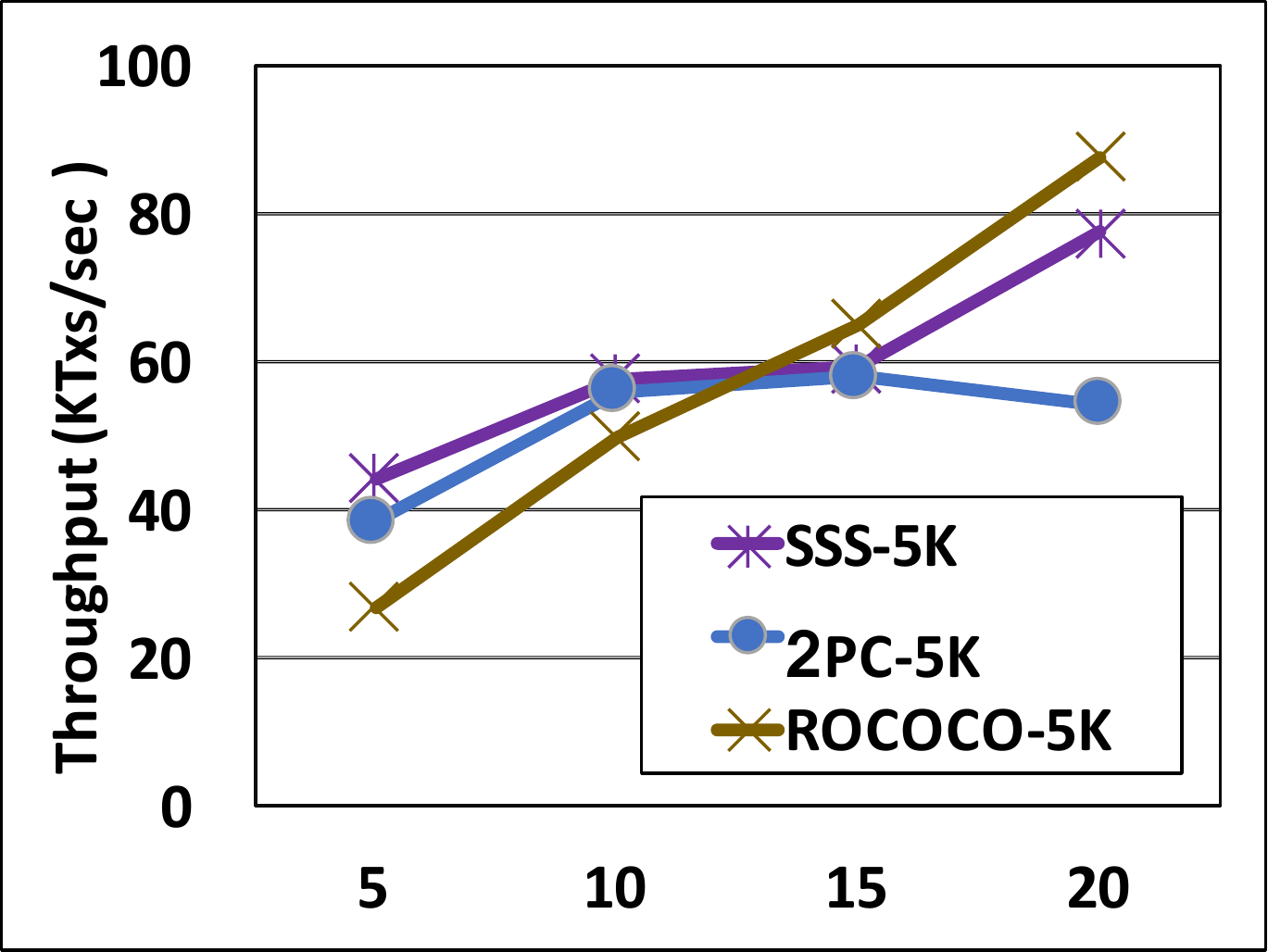}
    \label{fig:rococo-20}
}
\subfigure[80\%.]{
    \includegraphics[width=.2\textwidth]{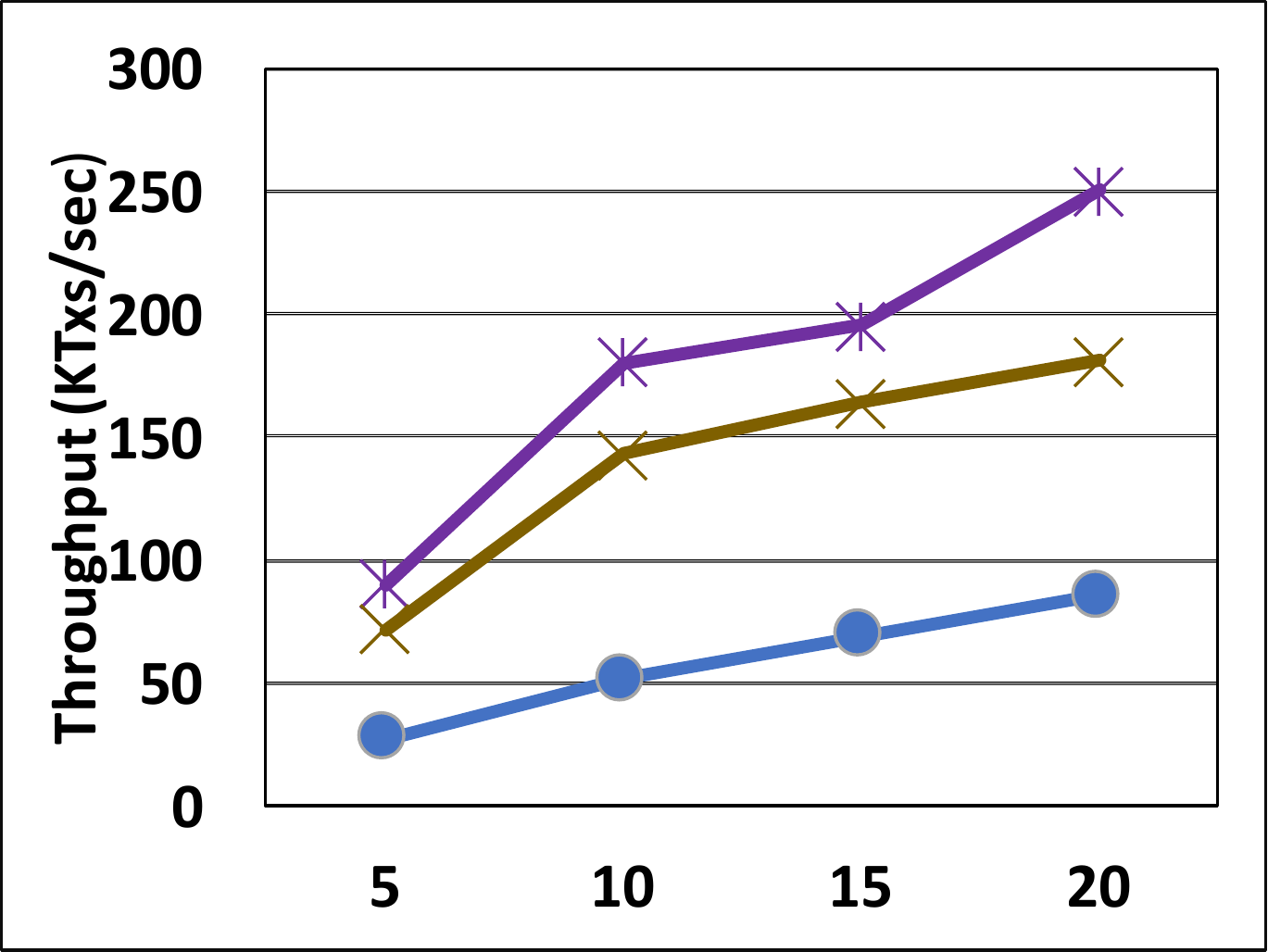}
    \label{fig:rococo-80}
}
\vspace{-5pt}

\caption{SSS, 2PC-baseline, ROCOCO varying \% of read-only transactions. Legend in (a) applies to (b).}
\label{yourass}
\end{figure}

In Figure~\ref{yourass} we compare SSS against ROCOCO and 2PC-baseline. To be compliant with ROCOCO, we disable replication for all competitors and we select 5k as total number of shared keys because ROCOCO finds its sweet spot in the presence of contention. Accesses are not local.

Figures~\ref{fig:rococo-20} and~\ref{fig:rococo-80} show the results with 20\% and 80\% read-only transactions respectively. In write intensive workload, ROCOCO slightly outperforms SSS due to its lock-free executions and its capability of re-ordering deferrable transaction pieces. However, even in this configuration, which matches a favorable scenario for ROCOCO, SSS is only 13\% slower than ROCOCO and 70\% faster than 2PC-baseline.
In read-intensive workload, SSS outperforms ROCOCO by 40\% and by almost 3$\times$ 2PC-baseline at 20 nodes. This gain is because ROCOCO is not optimized for read-only transactions; in fact, its read-only are not abort-free and they need to wait for all conflicting update transactions in order to execute.

We also configured the benchmark to produce 50\% of keys access locality, 
meaning the probability that a key is stored by the node where the transaction is executing (local node), and 50\% of uniform access.
Increasing local accesses has a direct impact on the application contention level. In fact, since each key is replicated on two nodes, remote communication is still needed by update transactions, while the number of objects accessible by a client reduces when the number of nodes increases (e.g., with 20 nodes and 5k keys, a client on a node can select its accessed keys among 250 keys rather than 5k). Read-only transactions are the ones that benefit the most from local accesses because they can leverage the local copy of each accessed key.

\begin{figure}[!htb]
   \begin{minipage}{0.24\textwidth}
     \centering
     \includegraphics[width=.92\textwidth]{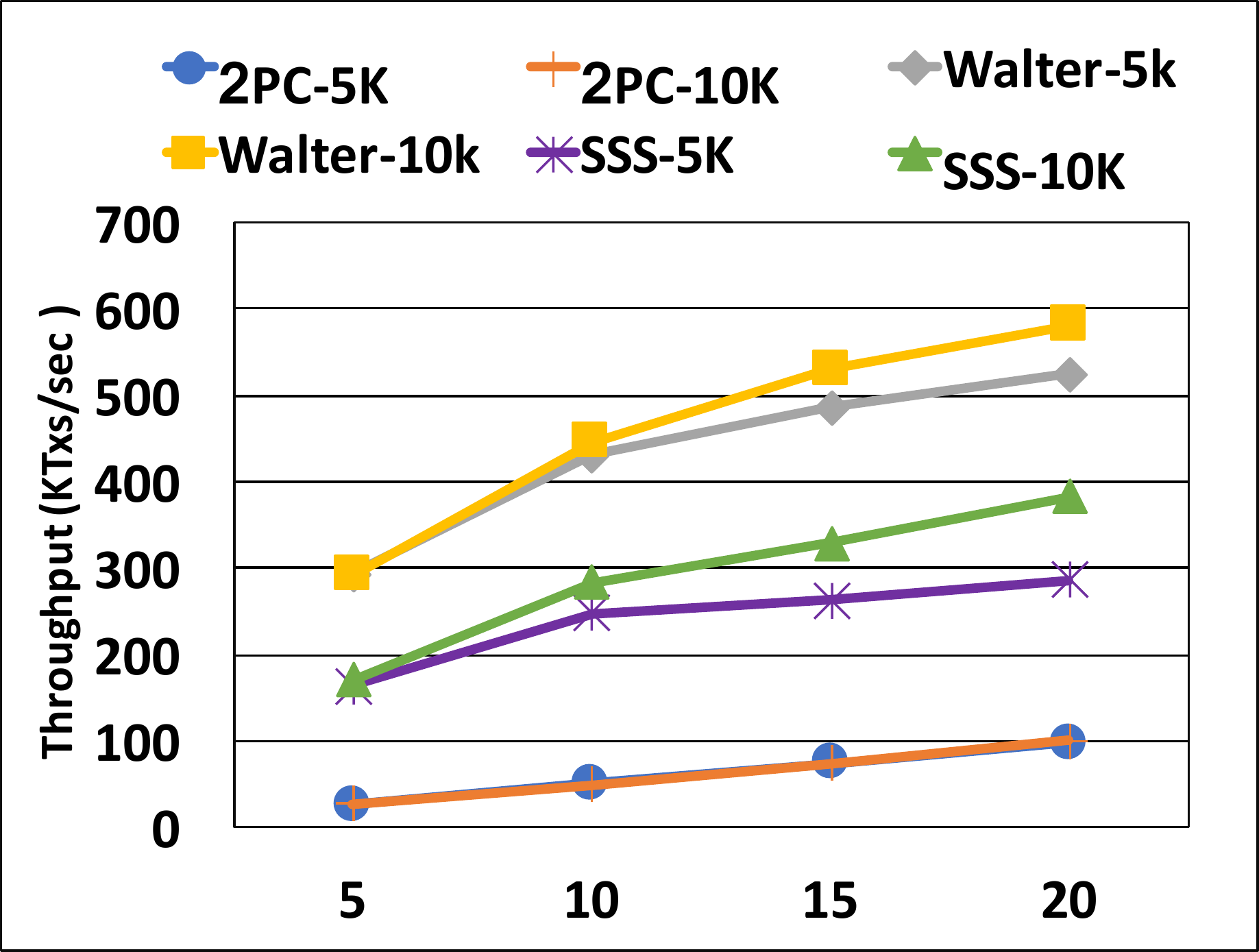}
\caption{Throughput 80\% read-only and 50\% locality.}
\label{local-stuff}
   \end{minipage}\hfill
   \begin{minipage}{0.24\textwidth}
     \centering
\includegraphics[width=1\textwidth]{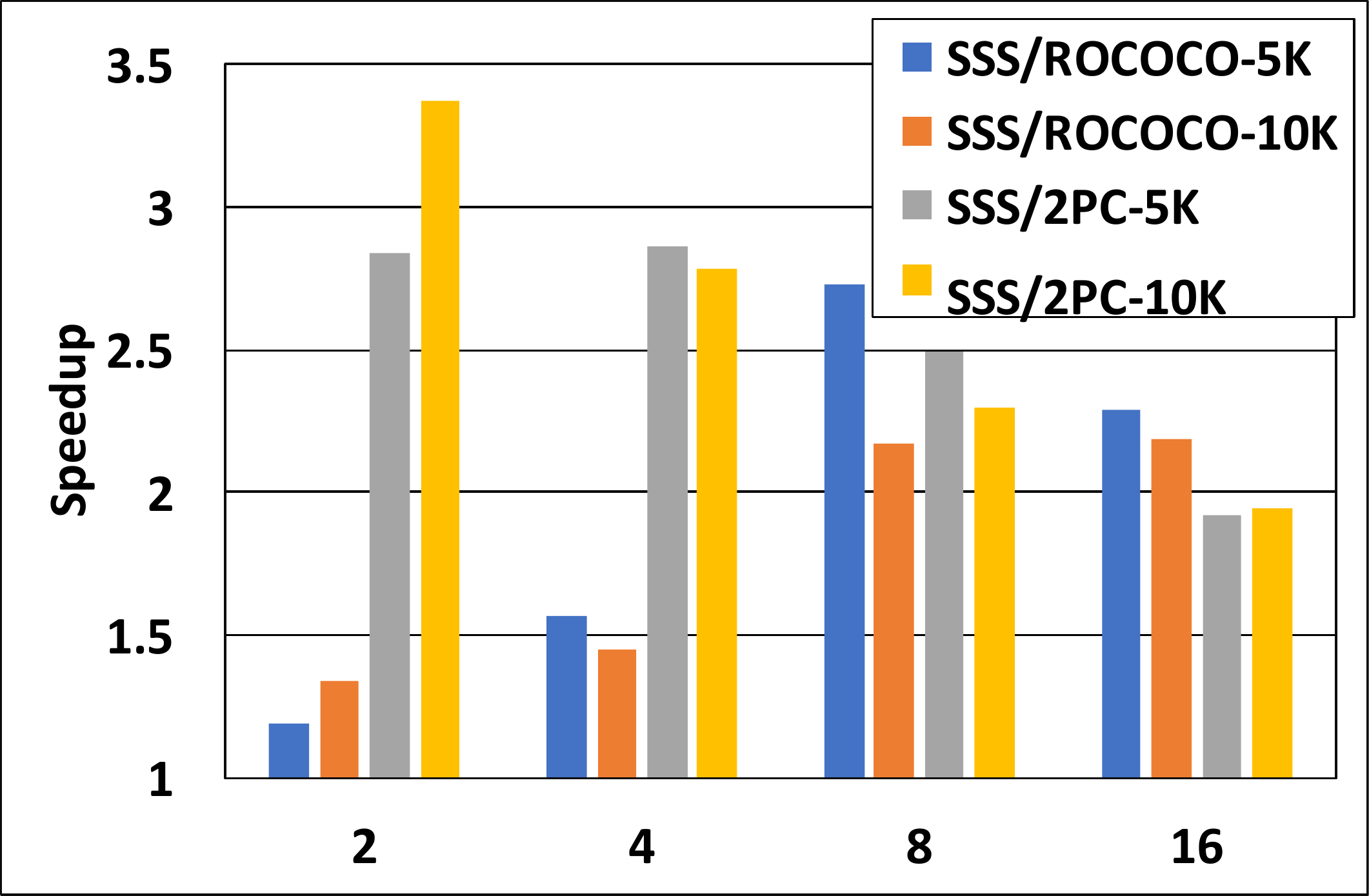}
\caption{Speedup of SSS over ROCOCO and 2PC-baseline increasing the size of read-only transactions.}
\label{bestplot}
   \end{minipage}
\end{figure}

We report the results (in Figure~\ref{local-stuff}) using the same configuration in Figure~\ref{fig:repl-thr-nolocal-c} because that is the most relevant to SSS and Walter.
Results confirm similar trend.
SSS is more than 3.5$\times$ faster than 2PC-baseline but, as opposed to the non-local case, here it cannot close the gap with Walter due to the high contention
around snapshot-queues.

In Figure~\ref{bestplot} we show the impact of increasing the number of read operations inside read-only transactions from 2 to 16. For this experiment we used 15 nodes and 80\% of read-only workload. Results report the ratio between the throughput of SSS and both ROCOCO and 2PC-baseline. When compared to ROCOCO, SSS shows a growing speedup, moving from 1.2$\times$ with 2 read operations to 2.2$\times$ with 16 read operations. This is because, as stated previously, ROCOCO encounters a growing number of aborts for read-only transactions while increasing accessed objects.
2PC-baseline degrades less than ROCOCO when operations increases because it needs less network communications for read-only transactions.

\section{Related Work}

Many distributed transactional repositories have been proposed in literature, examples include~\cite{granola,hstore,pnuts,dynamo,spanner,cure, gentlerain,rethinking,Cockroach}.
Among them, Spanner~\cite{spanner}, Scatter~\cite{scatter}, and ROCOCO~\cite{rococo} guarantee the same level of consistency as SSS. 

Google Spanner~\cite{spanner} is a high performance solution that leverages a global source of synchronization to timestamp transactions so that a total order among them can always be determined, including when nodes are in different geographic locations. This form of synchronization is materialized by the \textit{TrueTime} API. This API uses a combination of a very fast dedicated network, GPS, and atomic clocks to provide accuracy of the assigned timestamps.
Although outstanding, Spanner's architecture needs special-purpose hardware and therefore it cannot be easily adopted and extended.


Scatter provides external consistency on top of a Paxos-replicated log. The major difference with SSS is that Scatter only supports single key transactions while SSS provides a more general semantics.
ROCOCO uses a two-round protocol to establish an external schedule in the system, but it does not support abort-free read-only transactions.


Replicated Commit~\cite{rp} provides serializability by replicating the commit operation using 2PC in every data center and Paxos to establish consensus among data centers. As opposed to SSS, in Replicated Commit read operations require contacting all data centers and collect replies from a majority of them in order to proceed. SSS's read operations are handled by the fastest replying server.

Granola~\cite{granola} ensures serializability using a timestamp-based approach with a loosely synchronized clock per node. Granola provides its best performance when transactions can be defined as independent, meaning they can entirely execute on a single server.
SSS has no restriction on transaction accesses.

CockroachDB~\cite{Cockroach} uses a serializable optimistic concurrency control, which processes transactions by relying on multi-versioning and timestamp-ordering. The main difference with SSS is the way consistent reads are implemented. CockroachDB relies on consensus while SSS needs only to contact the fastest replica of an object.

Calvin~\cite{calvin}
uses a deterministic locking protocol supported by a sequencer layer that orders transactions. In order to do that, Calvin requires a priori knowledge on accessed read and written objects. Although the sequencer can potentially be able to assign transaction timestamp to meet external consistency requirements, SSS does that without assuming knowledge of read-set and write-set prior transaction execution and without the need of such a global source of synchronization.


SCORe~\cite{score}, guarantees similar properties as SSS, but it fails to ensure external consistency since it relies on a single non-synchronized scalar timestamp per node to order transactions, and therefore its abort-free read-only transactions might be forced to read old version of shared objects.

Other protocols, such as GMU~\cite{gmu12}, Walter~\cite{walter}, Clock-SI~\cite{clocksi} and Dynamo~\cite{dynamo},
provide scalability by supporting weaker levels of consistency.
GMU~\cite{gmu12} provides transactions with the possibility to read the latest version of an object by using vector clocks; however it cannot guarantee serializable transactions. Walter use a non-monotonic version of Snapshot Isolation (SI) that allows long state fork.
Clock-SI provides SI using a loosely synchronized clock scheme.


\section{Conclusions}
In this paper we presented SSS, a transactional repository that implements a novel distributed concurrency control providing external consistency without a global synchronization service. SSS is unique because it preserves the above properties while guaranteeing abort-free read-only transactions. The combination of snapshot-queuing and vector clock is the key technique that makes SSS possible. Results confirmed significant speedup over state-of-the-art competitors in read-dominated workloads.

\newpage
\bibliographystyle{abbrv}
\bibliography{bib/bibl}

\begin{thebibliography}{10}

\bibitem{clamson}
{CloudLab Clemson}, 2017.
\newblock \url{http://docs.cloudlab.us/hardware.html}.

\bibitem{adya}
A.~Adya.
\newblock {\em Weak Consistency: A Generalized Theory and Optimistic
  Implementations for Distributed Transactions}.
\newblock PhD thesis, 1999.
\newblock AAI0800775.

\bibitem{cure}
D.~D. Akkoorath, A.~Z. Tomsic, M.~Bravo, Z.~Li, T.~Crain, A.~Bieniusa,
  N.~Pregui{\c{c}}a, and M.~Shapiro.
\newblock Cure: Strong semantics meets high availability and low latency.
\newblock In {\em Distributed Computing Systems (ICDCS), 2016 IEEE 36th
  International Conference on}, pages 405--414. IEEE, 2016.

\bibitem{DBLP:conf/sigmetrics/AtikogluXFJP12}
B.~Atikoglu, Y.~Xu, E.~Frachtenberg, S.~Jiang, and M.~Paleczny.
\newblock Workload analysis of a large-scale key-value store.
\newblock In P.~G. Harrison, M.~F. Arlitt, and G.~Casale, editors, {\em {ACM}
  {SIGMETRICS/PERFORMANCE} Joint International Conference on Measurement and
  Modeling of Computer Systems, {SIGMETRICS} '12, London, United Kingdom, June
  11-15, 2012}, pages 53--64. {ACM}, 2012.

\bibitem{rethinking}
P.~A. Bernstein and S.~Das.
\newblock Rethinking eventual consistency.
\newblock In {\em Proceedings of the 2013 ACM SIGMOD International Conference
  on Management of Data}, SIGMOD '13, pages 923--928, New York, NY, USA, 2013.
  ACM.

\bibitem{serializability}
P.~A. Bernstein and N.~Goodman.
\newblock Concurrency control in distributed database systems.
\newblock {\em ACM Computing Surveys (CSUR)}, 13(2):185--221, 1981.

\bibitem{strict}
P.~A. Bernstein, V.~Hadzilacos, and N.~Goodman.
\newblock Concurrency control and recovery in database systems.
\newblock 1987.

\bibitem{Cockroach}
{Cockroach Labs}.
\newblock {CockroachDB }, 2017.
\newblock \url{https://github.com/cockroachdb/cockroach}.

\bibitem{pnuts}
B.~F. Cooper, R.~Ramakrishnan, U.~Srivastava, A.~Silberstein, P.~Bohannon,
  H.-A. Jacobsen, N.~Puz, D.~Weaver, and R.~Yerneni.
\newblock Pnuts: Yahoo!'s hosted data serving platform.
\newblock {\em Proceedings of the VLDB Endowment}, 1(2):1277--1288, 2008.

\bibitem{DBLP:conf/cloud/CooperSTRS10}
B.~F. Cooper, A.~Silberstein, E.~Tam, R.~Ramakrishnan, and R.~Sears.
\newblock Benchmarking cloud serving systems with {YCSB}.
\newblock In J.~M. Hellerstein, S.~Chaudhuri, and M.~Rosenblum, editors, {\em
  Proceedings of the 1st {ACM} Symposium on Cloud Computing, SoCC 2010,
  Indianapolis, Indiana, USA, June 10-11, 2010}, pages 143--154. {ACM}, 2010.

\bibitem{spanner}
J.~C. Corbett, J.~Dean, M.~Epstein, A.~Fikes, C.~Frost, J.~J. Furman,
  S.~Ghemawat, A.~Gubarev, C.~Heiser, P.~Hochschild, W.~Hsieh, S.~Kanthak,
  E.~Kogan, H.~Li, A.~Lloyd, S.~Melnik, D.~Mwaura, D.~Nagle, S.~Quinlan,
  R.~Rao, L.~Rolig, Y.~Saito, M.~Szymaniak, C.~Taylor, R.~Wang, and
  D.~Woodford.
\newblock {Spanner: Google's Globally Distributed Database}.
\newblock {\em ACM Trans. Comput. Syst.}, 31(3):8:1--8:22, Aug. 2013.

\bibitem{granola}
J.~A. Cowling and B.~Liskov.
\newblock Granola: Low-overhead distributed transaction coordination.
\newblock In {\em USENIX Annual Technical Conference}, volume~12, 2012.

\bibitem{dynamo}
G.~DeCandia, D.~Hastorun, M.~Jampani, G.~Kakulapati, A.~Lakshman, A.~Pilchin,
  S.~Sivasubramanian, P.~Vosshall, and W.~Vogels.
\newblock Dynamo: amazon's highly available key-value store.
\newblock In {\em ACM SIGOPS operating systems review}, volume~41, pages
  205--220. ACM, 2007.

\bibitem{DBLP:journals/csur/DefagoSU04}
X.~D{\'{e}}fago, A.~Schiper, and P.~Urb{\'{a}}n.
\newblock Total order broadcast and multicast algorithms: Taxonomy and survey.
\newblock {\em {ACM} Comput. Surv.}, 36(4):372--421, 2004.

\bibitem{clocksi}
J.~Du, S.~Elnikety, and W.~Zwaenepoel.
\newblock Clock-si: Snapshot isolation for partitioned data stores using
  loosely synchronized clocks.
\newblock In {\em Reliable Distributed Systems (SRDS), 2013 IEEE 32nd
  International Symposium on}, pages 173--184. IEEE, 2013.

\bibitem{gentlerain}
J.~Du, C.~Iorgulescu, A.~Roy, and W.~Zwaenepoel.
\newblock Gentlerain: Cheap and scalable causal consistency with physical
  clocks.
\newblock In {\em Proceedings of the ACM Symposium on Cloud Computing}, pages
  1--13. ACM, 2014.

\bibitem{exconsistency}
D.~K. Gifford.
\newblock {\em Information storage in a decentralized computer system}.
\newblock PhD thesis, Stanford University, 1981.

\bibitem{scatter}
L.~Glendenning, I.~Beschastnikh, A.~Krishnamurthy, and T.~Anderson.
\newblock Scalable consistency in scatter.
\newblock In {\em Proceedings of the Twenty-Third ACM Symposium on Operating
  Systems Principles}, pages 15--28. ACM, 2011.

\bibitem{2pc}
J.~Gray and L.~Lamport.
\newblock Consensus on transaction commit.
\newblock {\em ACM Transactions on Database Systems (TODS)}, 31(1):133--160,
  2006.

\bibitem{hstore}
R.~Kallman, H.~Kimura, J.~Natkins, A.~Pavlo, A.~Rasin, S.~Zdonik, E.~P. Jones,
  S.~Madden, M.~Stonebraker, Y.~Zhang, et~al.
\newblock H-store: a high-performance, distributed main memory transaction
  processing system.
\newblock {\em Proceedings of the VLDB Endowment}, 1(2):1496--1499, 2008.

\bibitem{VCOptimized}
T.~Landes.
\newblock Dynamic vector clocks for consistent ordering of events in dynamic
  distributed applications.
\newblock In {\em PDPTA}, pages 31--37, 2006.

\bibitem{rp}
H.~Mahmoud, F.~Nawab, A.~Pucher, D.~Agrawal, and A.~El~Abbadi.
\newblock Low-latency multi-datacenter databases using replicated commit.
\newblock {\em Proceedings of the VLDB Endowment}, 6(9):661--672, 2013.

\bibitem{rococo}
S.~Mu, Y.~Cui, Y.~Zhang, W.~Lloyd, and J.~Li.
\newblock Extracting more concurrency from distributed transactions.
\newblock In {\em OSDI}, volume~14, pages 479--494, 2014.

\bibitem{score}
S.~Peluso, P.~Romano, and F.~Quaglia.
\newblock {SCORe}: A scalable one-copy serializable partial replication
  protocol.
\newblock In {\em Middleware 2012}, pages 456--475, 2012.

\bibitem{gmu12}
S.~Peluso, P.~Ruivo, P.~Romano, F.~Quaglia, and L.~Rodrigues.
\newblock Gmu: Genuine multiversion update-serializable partial data
  replication.
\newblock {\em IEEE Transactions on Parallel and Distributed Systems},
  27(10):2911--2925, 2016.

\bibitem{ricci2014introducing}
R.~Ricci, E.~Eide, and C.~Team.
\newblock Introducing cloudlab: Scientific infrastructure for advancing cloud
  architectures and applications.
\newblock {\em ; login:: the magazine of USENIX \& SAGE}, 39(6):36--38, 2014.

\bibitem{walter}
Y.~Sovran, R.~Power, M.~K. Aguilera, and J.~Li.
\newblock Transactional storage for geo-replicated systems.
\newblock In {\em Proceedings of the Twenty-Third ACM Symposium on Operating
  Systems Principles}, pages 385--400. ACM, 2011.

\bibitem{calvin}
A.~Thomson, T.~Diamond, S.-C. Weng, K.~Ren, P.~Shao, and D.~J. Abadi.
\newblock Calvin: fast distributed transactions for partitioned database
  systems.
\newblock In {\em Proceedings of the 2012 ACM SIGMOD International Conference
  on Management of Data}, pages 1--12. ACM, 2012.

\bibitem{DBLP:conf/sosp/TuZKLM13}
S.~Tu, W.~Zheng, E.~Kohler, B.~Liskov, and S.~Madden.
\newblock Speedy transactions in multicore in-memory databases.
\newblock In M.~Kaminsky and M.~Dahlin, editors, {\em {ACM} {SIGOPS} 24th
  Symposium on Operating Systems Principles, {SOSP} '13, Farmington, PA, USA,
  November 3-6, 2013}, pages 18--32. {ACM}, 2013.

\bibitem{VCOptimized1}
X.~Wang, J.~Mayo, W.~Gao, and J.~Slusser.
\newblock An efficient implementation of vector clocks in dynamic systems.
\newblock In {\em PDPTA}, pages 593--599, 2006.

\bibitem{carousel}
X.~Yan, L.~Yang, H.~Zhang, X.~C. Lin, B.~Wong, K.~Salem, and T.~Brecht.
\newblock Carousel: Low-latency transaction processing for globally-distributed
  data.
\newblock In {\em Proceedings of the 2018 International Conference on
  Management of Data}, pages 231--243. ACM, 2018.

\bibitem{tapir}
I.~Zhang, N.~K. Sharma, A.~Szekeres, A.~Krishnamurthy, and D.~R. Ports.
\newblock Building consistent transactions with inconsistent replication.
\newblock In {\em Proceedings of the 25th Symposium on Operating Systems
  Principles}, pages 263--278. ACM, 2015.

\end{thebibliography}

\end{document}